\documentclass[twoside,11pt]{article}

\usepackage{tikz}
\usetikzlibrary{
  arrows.meta,
  positioning,
  shapes.geometric,
  shapes.symbols,
  fit,
  calc,
  backgrounds
}

\usepackage[preprint]{jmlr2e}

\usepackage[T1]{fontenc}
\usepackage[utf8]{inputenc}
\usepackage{amsmath}
\usepackage{amssymb}
\usepackage{amsfonts}
\usepackage{graphicx}
\graphicspath{{figures/}}
\usepackage{booktabs}
\usepackage{multirow}
\usepackage{float}
\usepackage{placeins}
\usepackage{algorithm}
\usepackage{algpseudocode}
\usepackage{url}
\usepackage{hyperref}

\begin{document}

\title{Theorem-Grounded Execution Ontologies for Interpretable Machine Reasoning}

\author{\name Raghu Anantharangachar \email rag.anant2011@gmail.com \\
       \addr Independent Researcher\\
       Bangalore, India}

\editor{}
\maketitle

\begin{abstract}

Large language models have demonstrated remarkable performance on reasoning tasks across mathematics, science, programming, and commonsense reasoning \citep{brown2020gpt3,openai2023gpt4}. Despite these advances, the internal reasoning processes of modern language models remain largely latent, making them difficult to interpret, verify, replay, and transfer across domains. Existing approaches such as chain-of-thought prompting \citep{wei2022chain}, tree-of-thoughts \citep{yao2023tree}, graph-of-thoughts \citep{besta2024graph}, and tool-augmented reasoning \citep{yao2023react,schick2023toolformer} expose intermediate reasoning artifacts but typically lack explicit execution semantics, formal state representations, and verifiable reasoning structures.

In this work, we introduce \emph{Theorem-Grounded Execution Ontologies (TGEO)}, a framework that represents reasoning as an executable state-transition process rather than a sequence of generated tokens. Given an input problem, the framework identifies relevant theorem families, binds the problem to a domain-specific ontology, discovers objects, instantiates states and operators, constructs predicates and contracts, and synthesizes an executable reasoning graph. The resulting execution graph provides an interpretable and replayable representation of the reasoning process in which every state transition, operator application, and validation step is explicitly represented.

The proposed architecture integrates five key components: (1) theorem-grounded reasoning priors, (2) executable ontologies, (3) operator-mediated state transitions, (4) predicate and contract-based execution validation, and (5) architectural auditing and failure localization. Together, these components transform reasoning into a structured execution process that can be inspected, verified, and reproduced.

We evaluate the framework on a collection of theorem-intensive reasoning tasks derived from mathematical benchmark domains \citep{hendrycks2021mmlu,cobbe2021gsm8k,hendrycks2021math} and a curated Golden Execution Suite. In addition to answer correctness, we introduce a comprehensive evaluation methodology that measures theorem assignment, ontology coverage, planner activation, operator utilization, state materialization, execution coverage, replayability, and architectural health. Experimental results demonstrate that theorem assignment and ontology construction achieve high reliability, while architectural auditing identifies state materialization and predicate satisfaction as the primary bottlenecks limiting end-to-end execution performance.

Our findings suggest that theorem-grounded execution ontologies provide a promising alternative to purely latent reasoning representations by enabling explicit, executable, and verifiable reasoning processes. More broadly, the proposed framework establishes a foundation for reusable reasoning structures that can support interpretable machine reasoning, cross-domain knowledge transfer, and future research into executable reasoning substrates for advanced intelligent systems.

\end{abstract}

\section{Introduction}
\label{sec:introduction}

Recent advances in large language models (LLMs) have produced substantial improvements on a wide range of reasoning benchmarks, including mathematical reasoning, scientific question answering, commonsense reasoning, and code generation \citep{brown2020gpt3,openai2023gpt4}. Models such as GPT-4, Claude, Gemini, and other frontier systems demonstrate the ability to solve complex tasks that were previously considered beyond the reach of language-based approaches. Despite these improvements, the internal reasoning processes of these systems remain largely opaque. The final answer is observable, but the intermediate reasoning structures that lead to the answer are typically represented only as latent activations within the model.

Several techniques have been proposed to improve the visibility of reasoning processes. Chain-of-thought prompting \citep{wei2022chain} encourages models to generate intermediate reasoning steps in natural language. Self-consistency \citep{wang2023selfconsistency} improves robustness by aggregating multiple reasoning paths. Tree-of-Thoughts \citep{yao2023tree} extends reasoning into structured search over alternative thought trajectories, while ReAct \citep{yao2023react} combines reasoning and action generation. More recently, Graph-of-Thoughts \citep{besta2024graph} has explored graph-based reasoning structures that capture richer dependencies among intermediate reasoning units.

Although these methods expose reasoning traces, the resulting artifacts are primarily textual. They do not provide explicit execution semantics, formally defined state transitions, operator contracts, or replayable execution structures. As a consequence, it remains difficult to verify whether a reasoning process is internally consistent, determine which operations were performed, identify where failures occurred, or reuse reasoning structures across domains.

Classical artificial intelligence approached reasoning differently. Planning systems \citep{strips1971,pddl1998} represented problems using explicit states, actions, preconditions, and effects. Knowledge representation systems employed ontologies, symbolic relations, and logical inference mechanisms \citep{gruber1993,guarino1998}. While these approaches provided strong interpretability and formal guarantees, they often lacked the flexibility and broad knowledge coverage demonstrated by modern language models.

This paper explores a middle ground between purely latent neural reasoning and purely symbolic reasoning. We introduce a framework for theorem-grounded execution ontologies that transforms reasoning tasks into explicit executable structures. Given a problem instance, the framework identifies an appropriate theorem family, binds the problem to a domain ontology, materializes domain-specific objects and states, selects executable operators, and constructs an execution graph that can be replayed and inspected. Reasoning is represented as a sequence of state transitions mediated by operators rather than as an unstructured sequence of generated tokens.

The proposed framework introduces five key ideas.

First, reasoning tasks are grounded in theorem families that provide an explicit bridge between problem statements and executable reasoning structures. Theorem assignment serves as a mechanism for selecting appropriate reasoning patterns before execution begins.

Second, domain ontologies are used to represent objects, states, operators, predicates, and contracts. Ontologies provide a structured representation of domain knowledge that can be reused across related problems.

Third, execution is represented as a state-transition system. Operators consume and produce typed states, allowing reasoning trajectories to be represented as executable graphs rather than textual traces.

Fourth, contracts and predicates are used to validate execution. Contracts specify preconditions and postconditions, while predicates encode semantic constraints that must hold throughout execution.

Fifth, reasoning traces are replayable. Every execution graph can be reconstructed and inspected, enabling detailed analysis of success and failure modes.

To evaluate the framework, we introduce a collection of reasoning tasks derived from mathematical benchmark domains and domain-specific reasoning examples. We measure theorem assignment, ontology coverage, operator utilization, execution replayability, execution graph validity, and architectural health. We further introduce execution-funnel diagnostics that localize failures across theorem assignment, ontology binding, planning, state transition, and goal achievement stages.

Our results demonstrate that theorem-grounded execution ontologies provide a practical mechanism for constructing interpretable reasoning structures. The framework achieves high theorem assignment rates, broad ontology coverage, replayable execution traces, and explicit reasoning representations. The experiments also reveal that state materialization and predicate satisfaction remain critical bottlenecks for large-scale executable reasoning systems.

The primary contribution of this work is not a new prompting strategy \citep{wei2022chain} or a new language model architecture \citep{brown2020gpt3,openai2023gpt4}. Instead, we present a framework that transforms reasoning into an explicit executable process whose intermediate structures can be inspected, validated, replayed, and potentially transferred across domains. We believe that such executable representations provide an important step toward more interpretable and verifiable reasoning systems.

\section{Related Work}
\label{sec:related_work}

The proposed theorem-grounded execution ontology framework lies at the intersection of several active research areas, including reasoning in large language models, automated theorem proving, planning, knowledge representation, neuro-symbolic learning, program synthesis, cognitive architectures, world models, and artificial general intelligence. This section reviews the most relevant literature and positions our contribution relative to these research directions.

\subsection{Reasoning in Large Language Models}

Recent advances in large language models (LLMs) have significantly improved performance on reasoning tasks across mathematics, science, commonsense reasoning, and code generation \citep{brown2020gpt3,openai2023gpt4}. Scaling studies demonstrated that increasingly large transformer models exhibit emergent reasoning capabilities that were not present in smaller systems \citep{brown2020gpt3}.

Chain-of-thought (CoT) prompting introduced the idea of explicitly generating intermediate reasoning steps before producing a final answer \citep{wei2022chain}. Subsequent work showed that CoT substantially improves performance on arithmetic, symbolic, and commonsense reasoning benchmarks \citep{wei2022chain}. Self-consistency decoding further improved reasoning performance by sampling multiple reasoning trajectories and selecting the most consistent answer \citep{wang2023selfconsistency}.

Additional prompting strategies have been proposed to improve reasoning. Zero-shot reasoning \citep{kojima2022largelanguagemodels} demonstrated that simple prompting instructions can elicit reasoning behavior. Least-to-most prompting decomposes difficult problems into sequences of simpler subproblems \citep{zhou2023leasttomost}. Program-aided reasoning \citep{gao2023pal} and Program-of-Thought prompting \citep{chen2023programofthoughts} incorporate executable programs into reasoning workflows.

Despite their effectiveness, these approaches remain largely textual. Intermediate reasoning steps are represented as natural language sequences rather than formally executable structures. Consequently, reasoning traces are difficult to verify, replay, or analyze systematically.

\subsection{Structured Reasoning and Search-Based Inference}

Several methods extend chain-of-thought reasoning \citep{wei2022chain} through explicit search.

Tree-of-Thoughts (ToT) formulates reasoning as search over alternative reasoning trajectories \citep{yao2023tree}. Rather than committing to a single reasoning path, ToT explores multiple candidate reasoning states and evaluates their quality before proceeding.

Graph-of-Thoughts (GoT) generalizes this idea by representing reasoning as a graph rather than a tree \citep{besta2024graph}. This allows information sharing across reasoning paths and supports more complex dependency structures.

Other approaches include deliberative decoding, Monte Carlo search, self-refinement methods, and multi-agent reasoning systems. While these methods improve exploration and planning, they continue to operate primarily over textual reasoning representations.

In contrast, our framework introduces explicit execution semantics through states, operators, predicates, contracts, and execution graphs.

\subsection{Tool Use and Agent Architectures}

The integration of reasoning with external actions has become an important area of research.

ReAct combines reasoning and action generation by interleaving thought generation with environment interaction \citep{yao2023react}. Toolformer extends this idea by enabling language models to learn tool usage automatically \citep{schick2023toolformer}.

Subsequent work has explored web browsing agents, code execution agents, autonomous task execution systems, and planner-based agent architectures \citep{yao2022webgpt}.

Modern agent systems often employ planning, decomposition, memory, retrieval, and tool invocation. Examples include planner-executor architectures, LangGraph-based workflows, AutoGPT-style systems, and multi-agent coordination frameworks.

These systems introduce explicit actions but typically lack formal state-transition semantics. Actions are represented procedurally rather than through reusable ontological structures. The present work can be viewed as a formalization of agent reasoning in which operators, predicates, contracts, and state transitions become first-class reasoning objects.

\subsection{Automated Theorem Proving}

Automated theorem proving (ATP) \citep{harrison2009logic,robinson1965resolution} represents one of the longest-standing research areas in artificial intelligence.

Classical theorem provers such as resolution-based systems demonstrated that logical reasoning can be represented through explicit symbolic structures \citep{strips1971}. Interactive theorem provers and proof assistants subsequently provided powerful environments for constructing machine-verifiable proofs.

Modern theorem proving systems include Lean \citep{lean2020}, Coq \citep{coq2021}, Isabelle/HOL \citep{isabelle2021}, HOL Light, and Metamath. These systems provide rich formal languages for representing mathematical knowledge and constructing proofs.

Recent work has increasingly incorporated machine learning into theorem proving \citep{rocktaschel2017endtoend,alphaproof2024}. Neural-guided theorem provers \citep{rocktaschel2017endtoend} and reinforcement-learning-based proof search systems have demonstrated promising results. Large language models have also been applied to formal proof generation and proof completion.

AlphaProof represents a recent large-scale effort combining machine learning and formal theorem proving \citep{alphaproof2024}.

Unlike formal proof systems \citep{lean2020,coq2021,isabelle2021}, the present work uses theorem families as reasoning priors that guide ontology selection and execution planning. The objective is not formal proof generation but the construction of executable reasoning structures.

\subsection{Knowledge Representation and Ontologies}

Knowledge representation has long been a foundational area of artificial intelligence.

Semantic networks, frames, description logics, and ontologies provide mechanisms for representing entities, relations, constraints, and domain knowledge \citep{gruber1993,guarino1998,hogan2021kg,ji2021kg}.

Large-scale ontology systems such as Cyc \citep{cyc1989}, WordNet \citep{miller1995wordnet}, ConceptNet \citep{speer2017conceptnet}, Freebase \citep{bollacker2008freebase}, DBpedia, and Wikidata \citep{vrandevcic2014wikidata} demonstrated the value of structured knowledge representations.

Ontologies have been widely used in biomedical informatics, scientific reasoning, semantic web systems, and intelligent agents \citep{guarino1998,hogan2021kg}.

Most ontology systems focus on static knowledge representation. The proposed framework extends this perspective by introducing executable semantics through states, operators, predicates, contracts, and execution graphs.

\subsection{Knowledge Graph Completion and Graph Reasoning}

Knowledge graph completion \citep{hogan2021kg,ji2021kg} seeks to infer missing relationships among entities in large relational graphs.

Early approaches such as TransE \citep{bordes2013transe} introduced translational embedding methods for representing multi-relational data. Subsequent work developed more expressive embedding models, including ConvE \citep{dettmers2018conve}, tensor factorization methods, and graph neural network approaches.

Relational graph convolutional networks (R-GCNs) extended graph neural networks to relational data \citep{schlichtkrull2018rgcn}. More recently, graph transformers and graph retrieval systems have demonstrated strong performance across knowledge-intensive tasks \citep{ji2021kg}.

Knowledge graph reasoning systems primarily focus on static relationships \citep{hogan2021kg}. In contrast, our framework represents reasoning as executable transitions among dynamic states.

\subsection{Neuro-Symbolic Learning and Reasoning}

Neuro-symbolic reasoning \citep{garcez2019neurosymbolic} seeks to combine the representational power of neural networks with the interpretability and compositionality of symbolic systems.

Early neuro-symbolic systems explored the integration of logic and neural computation \citep{garcez2002neuralsymbolic}. More recent work has investigated neural theorem proving, differentiable logic, symbolic rule induction, and hybrid reasoning architectures \citep{garcez2019neurosymbolic}.

A recurring theme within neuro-symbolic research \citep{garcez2019neurosymbolic} is the need for representations that support both learning and reasoning. The proposed framework shares this objective by combining learned theorem assignment and ontology induction with explicit symbolic execution structures.

Unlike many neuro-symbolic approaches that focus on logical inference \citep{garcez2019,deepproblog2018}, our framework emphasizes executable reasoning processes and replayable execution traces.

\subsection{Program Synthesis and Program Induction}

Program synthesis aims to automatically generate executable programs from specifications, examples, or natural language descriptions.

Recent advances in neural code generation have enabled language models to synthesize increasingly complex programs. Program-aided reasoning \citep{gao2023pal} and Program-of-Thought prompting \citep{chen2023programofthoughts} demonstrated that executable programs can improve reasoning performance.

Related work includes neural program synthesis, symbolic execution, differentiable interpreters, and program induction systems.

Execution graphs share several properties with synthesized programs. Both represent structured sequences of executable operations. However, execution graphs are derived from theorem-grounded ontologies rather than conventional programming languages.

\subsection{Planning and Decision Making}

Planning systems \citep{strips1971,pddl1998} represent problems through states, actions, preconditions, and effects.

STRIPS \citep{strips1971} established many foundational ideas in automated planning. The Planning Domain Definition Language (PDDL) later provided a standardized representation for planning domains \citep{pddl1998}.

Classical planning algorithms \citep{strips1971,pddl1998} provide strong guarantees regarding correctness and optimality but generally require manually specified domain models.

Model-based reinforcement learning, hierarchical planning, and world-model approaches extend planning concepts into learned environments.

The proposed framework inherits several ideas from planning research, including explicit state spaces, operators, and goal-directed execution. However, unlike classical planners, the framework attempts to induce these structures automatically from reasoning tasks.

\subsection{Cognitive Architectures}

Cognitive architectures seek to model general cognition through explicit computational structures.

Soar \citep{laird1987soar} introduced production-rule reasoning and goal-directed execution mechanisms. ACT-R \citep{actr1998} modeled cognition through interactions among memory systems, procedural rules, and attention mechanisms.

More recent architectures such as Sigma and CLARION continue this tradition.

The state-operator formulation adopted in the present work is closely aligned with several principles found in cognitive architectures, including explicit state representations, procedural execution, and goal-directed behavior.

\subsection{World Models and Structured Intelligence}

World models \citep{ha2018worldmodels,hafner2023dreamerv3} seek to learn representations that support prediction, planning, and decision making.

Early work demonstrated that latent models of environment dynamics can support planning and control \citep{ha2018worldmodels}. More recent systems such as DreamerV3 have demonstrated strong performance across diverse environments using learned world models \citep{hafner2023dreamerv3}.

LeCun's vision for autonomous machine intelligence similarly emphasizes predictive world models, planning, and hierarchical reasoning \citep{lecun2022path}.

The proposed framework can be interpreted as a reasoning-oriented world model in which theorem families, ontologies, states, and operators provide an explicit substrate for reasoning.

\subsection{Artificial General Intelligence}

Research on artificial general intelligence (AGI) emphasizes abstraction, transfer, compositionality, and generalization across domains.

Cognitive architectures, universal learning systems, and foundation models have all been proposed as potential paths toward general intelligence.

Recent discussions of AGI capabilities have highlighted the importance of reasoning, planning, tool use, and world modeling \citep{bubeck2023sparks}. Chollet argues that intelligence should be measured through generalization and adaptation rather than benchmark performance alone \citep{chollet2019arc}.

A recurring challenge in AGI research is the construction of reusable reasoning structures that transfer across domains. We view theorem-grounded execution ontologies as a step toward this objective by providing explicit representations of objects, states, operators, predicates, contracts, and execution graphs.

\subsection{Reasoning Benchmarks}

Reasoning benchmarks \citep{hendrycks2021mmlu} have played an important role in evaluating progress in machine reasoning.

MMLU \citep{hendrycks2021mmlu} provides broad coverage across academic disciplines and remains one of the most widely used benchmarks for evaluating language model reasoning capabilities.

Additional benchmarks include GSM8K \citep{cobbe2021gsm8k}, MATH \citep{hendrycks2021math}, ARC \citep{chollet2019arc}, GPQA \citep{rein2023gpqa}, theorem proving benchmarks \citep{alphaproof2024,lean2020}, code reasoning benchmarks \citep{chen2021codex,austin2021program}, and scientific reasoning datasets \citep{hendrycks2021mmlu}.

Most existing benchmarks \citep{hendrycks2021mmlu,cobbe2021gsm8k,hendrycks2021math,chollet2019arc} evaluate answer correctness without examining internal reasoning structures. The evaluation methodology introduced in this paper complements these benchmarks by measuring theorem assignment, ontology induction, planning, execution coverage, replayability, and architectural health.

\subsection{Summary}

Existing research has demonstrated substantial progress in reasoning, theorem proving, planning, program synthesis, knowledge representation, and neuro-symbolic learning \citep{garcez2019neurosymbolic,hogan2021kg,strips1971}. However, most modern reasoning systems continue to rely on latent representations or textual reasoning traces.

The proposed framework contributes a complementary perspective in which reasoning is represented as an explicit executable process grounded in theorem families, ontologies, states, operators, predicates, contracts, and execution graphs. By combining learned reasoning structures with executable semantics, the framework seeks to bridge the gap between modern language-model reasoning and classical symbolic execution systems.

\section{Problem Formulation}
\label{sec:problem_formulation}

\subsection{Motivation}

Contemporary reasoning systems typically represent reasoning implicitly through neural activations or textual reasoning traces. Although these approaches often produce correct answers, they provide limited insight into the underlying reasoning process. Intermediate reasoning structures are difficult to verify, replay, or transfer across domains.

We consider an alternative formulation in which reasoning is represented as an executable process operating over explicit domain structures. Rather than generating reasoning traces directly in natural language, the system constructs a theorem-grounded execution ontology consisting of objects, states, operators, predicates, contracts, and execution graphs.

The central problem addressed in this work is:

\begin{quote}
Given a reasoning problem, can a system automatically construct an executable reasoning representation that is interpretable, replayable, and capable of supporting formal execution?
\end{quote}

\subsection{Reasoning Task}

Let

\begin{equation}
x \in \mathcal{X}
\end{equation}

denote an input reasoning problem.

Examples include:

\begin{itemize}
    \item Mathematical reasoning tasks
    \item Scientific reasoning tasks
    \item Clinical decision-making problems
    \item Cybersecurity investigations
    \item Legal reasoning tasks
\end{itemize}

The objective is to produce

\begin{equation}
y \in \mathcal{Y}
\end{equation}

representing the final answer.

Traditional language models learn a direct mapping

\begin{equation}
f:\mathcal{X}\rightarrow\mathcal{Y}.
\end{equation}

In contrast, we seek to learn an explicit intermediate reasoning representation

\begin{equation}
f:\mathcal{X}
\rightarrow
(T,O,S,A,G)
\rightarrow
\mathcal{Y},
\end{equation}

where:

\begin{itemize}
    \item $T$ denotes theorem assignments,
    \item $O$ denotes ontology structures,
    \item $S$ denotes states,
    \item $A$ denotes operators,
    \item $G$ denotes execution graphs.
\end{itemize}

\subsection{Theorem-Grounded Reasoning}

We assume that every reasoning task belongs to one or more theorem families.

Let

\begin{equation}
\mathcal{T}
=
\{t_1,t_2,\ldots,t_n\}
\end{equation}

denote the set of theorem families.

Examples include:

\begin{equation}
\texttt{GroupTheory}
\end{equation}

\begin{equation}
\texttt{FieldTheory}
\end{equation}

\begin{equation}
\texttt{BayesianInference}
\end{equation}

\begin{equation}
\texttt{ClinicalDiagnosis}.
\end{equation}

A theorem assignment function

\begin{equation}
\phi_T:
\mathcal{X}
\rightarrow
\mathcal{T}
\end{equation}

maps an input problem to one or more theorem families.

The theorem acts as an executable reasoning prior that constrains subsequent ontology and operator selection.

\subsection{Execution Ontologies}

Each theorem family is associated with one or more ontologies.

Let

\begin{equation}
\mathcal{O}
=
\{o_1,o_2,\ldots,o_m\}
\end{equation}

denote the ontology space.

An ontology is defined as

\begin{equation}
o=
(\mathcal{V},
\mathcal{S},
\mathcal{A},
\mathcal{P},
\mathcal{C}),
\end{equation}

where

\begin{itemize}
    \item $\mathcal{V}$ is the object vocabulary,
    \item $\mathcal{S}$ is the state space,
    \item $\mathcal{A}$ is the operator space,
    \item $\mathcal{P}$ is the predicate set,
    \item $\mathcal{C}$ is the contract set.
\end{itemize}

Ontology assignment is defined as

\begin{equation}
\phi_O:
(T,x)
\rightarrow
O,
\end{equation}

which selects an executable ontology conditioned on both the problem and the theorem family.

\subsection{Objects}

Objects represent domain entities.

Let

\begin{equation}
v_i \in \mathcal{V}
\end{equation}

denote an object.

Examples include:

\paragraph{Mathematics}

\begin{equation}
\texttt{Field},
\texttt{Group},
\texttt{Subgroup}
\end{equation}

\paragraph{Healthcare}

\begin{equation}
\texttt{Patient},
\texttt{Diagnosis},
\texttt{Medication}
\end{equation}

\paragraph{Cybersecurity}

\begin{equation}
\texttt{Host},
\texttt{Credential},
\texttt{AttackPath}
\end{equation}

Object discovery extracts

\begin{equation}
V_x
=
\{v_1,\ldots,v_k\}
\end{equation}

for a given problem instance.

\subsection{States}

States represent executable configurations of objects.

A state is defined as

\begin{equation}
s=(v,\alpha),
\end{equation}

where

\begin{itemize}
    \item $v$ denotes an object,
    \item $\alpha$ denotes a set of properties.
\end{itemize}

The state space is

\begin{equation}
\mathcal{S}
=
\{s_1,s_2,\ldots\}.
\end{equation}

Examples include:

\begin{equation}
\texttt{SubgroupKnown}
\end{equation}

\begin{equation}
\texttt{DiagnosisConfirmed}
\end{equation}

\begin{equation}
\texttt{CredentialCompromised}.
\end{equation}

Reasoning proceeds through state transitions.

\subsection{Operators}

Operators represent executable reasoning actions.

An operator

\begin{equation}
a \in \mathcal{A}
\end{equation}

is defined as

\begin{equation}
a=
(\mathrm{pre}(a),
\mathrm{eff}(a)),
\end{equation}

where

\begin{itemize}
    \item $\mathrm{pre}(a)$ denotes preconditions,
    \item $\mathrm{eff}(a)$ denotes effects.
\end{itemize}

Examples include:

\paragraph{Mathematics}

\begin{equation}
\texttt{ComputeIndex}
\end{equation}

\begin{equation}
\texttt{ApplyLagrange}
\end{equation}

\paragraph{Healthcare}

\begin{equation}
\texttt{ConfirmDiagnosis}
\end{equation}

\paragraph{Cybersecurity}

\begin{equation}
\texttt{EscalatePrivilege}
\end{equation}

Operators induce state transitions

\begin{equation}
a:s_i \rightarrow s_j.
\end{equation}

\subsection{Predicates}

Predicates encode semantic conditions that must hold during execution.

Let

\begin{equation}
p:\mathcal{S}
\rightarrow
\{0,1\}
\end{equation}

denote a predicate.

Examples include:

\begin{equation}
\texttt{ValidSubgroup}
\end{equation}

\begin{equation}
\texttt{DiagnosisSupported}
\end{equation}

\begin{equation}
\texttt{CredentialExists}.
\end{equation}

Predicates determine operator applicability and execution validity.

\subsection{Contracts}

Contracts define correctness conditions for execution.

A contract is represented as

\begin{equation}
c=
(P_{\mathrm{pre}},
P_{\mathrm{post}}),
\end{equation}

where

\begin{itemize}
    \item $P_{\mathrm{pre}}$ specifies required predicates before execution,
    \item $P_{\mathrm{post}}$ specifies predicates that must hold after execution.
\end{itemize}

Contracts provide explicit verification semantics.

\subsection{Execution Graphs}

Reasoning is represented as an execution graph.

An execution graph is defined as

\begin{equation}
G=(N,E),
\end{equation}

where

\begin{equation}
N=
\{s_1,s_2,\ldots,s_n\}
\end{equation}

is the set of states and

\begin{equation}
E=
\{(s_i,a,s_j)\}
\end{equation}

is the set of operator-mediated transitions.

The graph defines a complete executable reasoning trace.

\subsection{Goal States}

Each reasoning problem specifies a goal condition

\begin{equation}
g \in \mathcal{S}.
\end{equation}

Execution seeks to construct a path

\begin{equation}
\pi=
(a_1,a_2,\ldots,a_k)
\end{equation}

such that

\begin{equation}
s_0
\xrightarrow{a_1}
s_1
\xrightarrow{a_2}
\cdots
\xrightarrow{a_k}
g,
\end{equation}

while satisfying all predicates and contracts.

\subsection{Learning Objective}

Given a dataset

\begin{equation}
D=
\{(x_i,y_i)\}_{i=1}^{N},
\end{equation}

the objective is to learn:

\begin{enumerate}
    \item Theorem assignment:
    \[
    \phi_T
    \]

    \item Ontology assignment:
    \[
    \phi_O
    \]

    \item State discovery:
    \[
    \phi_S
    \]

    \item Operator discovery:
    \[
    \phi_A
    \]

    \item Execution graph construction:
    \[
    \phi_G
    \]
\end{enumerate}

such that the resulting execution graph:

\begin{itemize}
    \item Achieves the correct answer,
    \item Satisfies contracts,
    \item Satisfies predicates,
    \item Is replayable,
    \item Remains interpretable.
\end{itemize}

We define the optimization objective as

\begin{equation}
\max \mathcal{L}
=
\lambda_1 \mathcal{L}_{\text{answer}}
+
\lambda_2 \mathcal{L}_{\text{execution}}
+
\lambda_3 \mathcal{L}_{\text{replay}}
+
\lambda_4 \mathcal{L}_{\text{interpretability}},
\end{equation}

where:

\begin{itemize}
    \item $\mathcal{L}_{\text{answer}}$ measures answer correctness,
    \item $\mathcal{L}_{\text{execution}}$ measures execution success,
    \item $\mathcal{L}_{\text{replay}}$ measures replayability,
    \item $\mathcal{L}_{\text{interpretability}}$ measures execution graph quality.
\end{itemize}

\subsection{Architectural Perspective}

The overall reasoning process is summarized as

\begin{equation}
x
\rightarrow
T
\rightarrow
O
\rightarrow
V
\rightarrow
S
\rightarrow
A
\rightarrow
G
\rightarrow
y.
\end{equation}

\begin{figure}[t]
\centering
\begin{tikzpicture}[
node distance=0.55cm,
stage/.style={
draw,
rectangle,
rounded corners,
minimum width=3.6cm,
minimum height=0.75cm,
align=center,
font=\small
},
arr/.style={-{Stealth[length=2mm]}, thick},
]

\node[stage] (n0) {Problem};
\node[stage,below=of n0] (n1) {Theorem};
\draw[arr] (n0) -- (n1);
\node[stage,below=of n1] (n2) {Ontology};
\draw[arr] (n1) -- (n2);
\node[stage,below=of n2] (n3) {Objects};
\draw[arr] (n2) -- (n3);
\node[stage,below=of n3] (n4) {States};
\draw[arr] (n3) -- (n4);
\node[stage,below=of n4] (n5) {Operators};
\draw[arr] (n4) -- (n5);
\node[stage,below=of n5] (n6) {Execution Graph};
\draw[arr] (n5) -- (n6);
\node[stage,below=of n6] (n7) {Answer};
\draw[arr] (n6) -- (n7);

\end{tikzpicture}
\caption{End-to-end theorem-grounded execution pipeline.}
\label{fig:problem_execution_pipeline}
\end{figure}
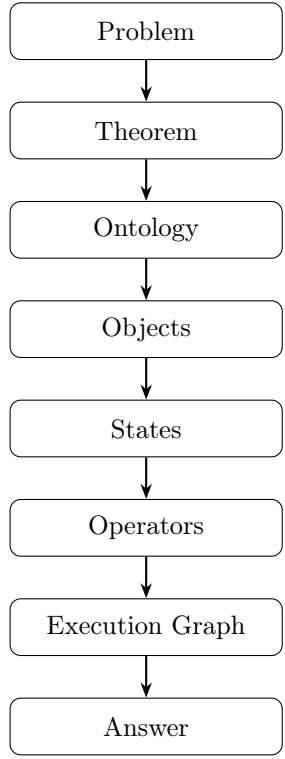

Equivalently,

Figure~\ref{fig:problem_execution_pipeline} summarizes the end-to-end pipeline.
% Auto-generated by scripts/figures/generate_problem_execution_pipeline.py

This formulation transforms reasoning from a latent sequence-generation process into an explicit executable state-transition system that can be inspected, replayed, verified, and analyzed.

\section{Theorem-Grounded Ontology Framework}
\label{sec:theorem_grounded_framework}

\subsection{Overview}

The central hypothesis of this work is that reasoning tasks can be represented through executable domain structures grounded in theorem families. Rather than directly generating reasoning traces, the proposed framework first identifies the theorem family governing a problem instance and then instantiates an executable ontology containing objects, states, operators, predicates, and contracts.

\begin{figure*}[t]
\centering
\begin{tikzpicture}[
node distance=1cm,
box/.style={
rectangle,
draw,
rounded corners,
minimum width=2cm,
minimum height=1.9cm,
align=center
},
>=Stealth
]

\node[box] (input) {Input \\Problem};

\node[box,right=of input] (theorem)
{Theorem \\Assignment};

\node[box,right=of theorem] (ontology)
{Ontology \\Grounding};

\node[box,right=of ontology] (planner)
{Planner \\Construction};

\node[box,right=of planner] (execution)
{Execution \\Graph};

\node[box,below=of planner]
(state)
{State \\Engine};

\node[box,below=of ontology]
(predicate)
{Predicate \\Validation};

\draw[->] (input) -- (theorem);
\draw[->] (theorem) -- (ontology);
\draw[->] (ontology) -- (planner);
\draw[->] (planner) -- (execution);

\draw[->] (planner) -- (state);
\draw[->] (state) -- (execution);

\draw[->] (predicate) -- (planner);
\draw[->] (ontology) -- (predicate);

\end{tikzpicture}
\caption{Architecture of the Theorem-Grounded Execution Ontology framework.}
\label{fig:tgeo_architecture}
\end{figure*}
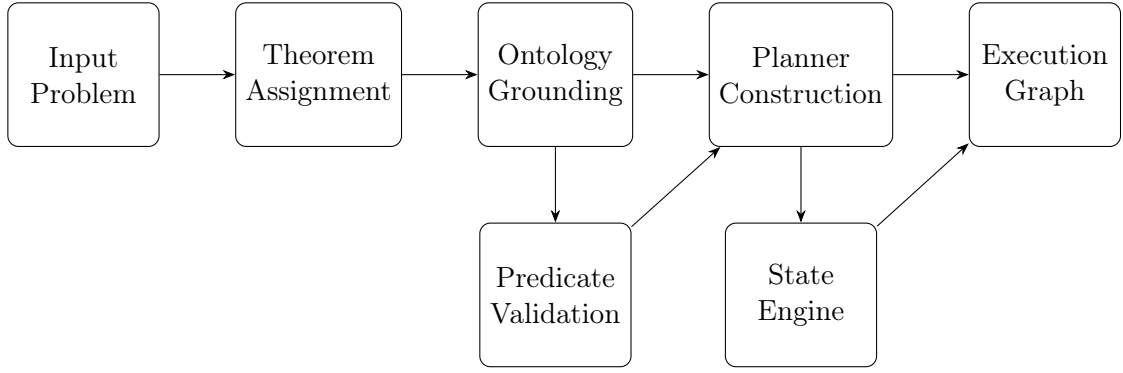

Figure~\ref{fig:tgeo_architecture} illustrates the overall framework.
% Auto-generated by scripts/figures/generate_tgeo_architecture.py

\begin{equation}
x
\rightarrow
T
\rightarrow
O
\rightarrow
V
\rightarrow
S
\rightarrow
A
\rightarrow
G
\rightarrow
y
\end{equation}

where:

\begin{itemize}
\item $x$ denotes the input problem,
\item $T$ denotes theorem assignments,
\item $O$ denotes ontology assignments,
\item $V$ denotes discovered objects,
\item $S$ denotes instantiated states,
\item $A$ denotes executable operators,
\item $G$ denotes the execution graph,
\item $y$ denotes the final answer.
\end{itemize}

This decomposition transforms reasoning from a latent generation process into an explicit executable reasoning pipeline.

\subsection{Theorem Assignment}

\begin{figure}[t]
\centering
\begin{tikzpicture}[
node distance=1.5cm,
box/.style={
draw,
rectangle,
rounded corners,
minimum width=2.2cm,
minimum height=0.8cm,
align=center
}
]

\node[box] (problem) {Problem};

\node[box,right=of problem]
(candidate) {Candidate\\Theorems};

\node[box,right=of candidate]
(rank) {Ranking};

\node[box,right=of rank]
(assign) {Assigned\\Theorem};

\draw[->] (problem)--(candidate);
\draw[->] (candidate)--(rank);
\draw[->] (rank)--(assign);

\end{tikzpicture}

\caption{Theorem assignment workflow.}
\label{fig:theorem_pipeline}
\end{figure}
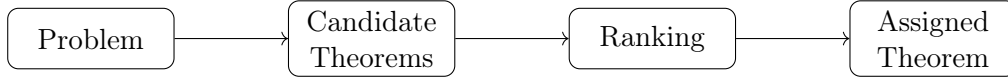

Figure~\ref{fig:theorem_pipeline} illustrates the theorem assignment pipeline.
% Auto-generated by scripts/figures/generate_theorem_pipeline.py

The first stage of the framework identifies the theorem family associated with a problem instance.

Let

\begin{equation}
\mathcal{T}
=
\{t_1,t_2,\ldots,t_n\}
\end{equation}

denote the theorem space.

Examples include:

\begin{itemize}
\item Group Theory
\item Ring Theory
\item Field Theory
\item Linear Algebra
\item Bayesian Inference
\item Clinical Diagnosis
\item Threat Analysis
\end{itemize}

The theorem assignment function is defined as

\begin{equation}
\phi_T :
\mathcal{X}
\rightarrow
2^{\mathcal{T}}
\end{equation}

where

\begin{equation}
T_x
=
\phi_T(x)
\end{equation}

represents the set of theorem families assigned to problem $x$.

The assignment process provides an explicit reasoning prior that constrains ontology selection and subsequent execution.

\subsection{Ontology Selection}

Each theorem family is associated with one or more executable ontologies \citep{gruber1993,guarino1998}.

Let

\begin{equation}
\mathcal{O}
=
\{o_1,o_2,\ldots,o_m\}
\end{equation}

denote the ontology space.

Ontology assignment is defined as

\begin{equation}
\phi_O :
(T_x,x)
\rightarrow
O_x.
\end{equation}

The selected ontology provides the vocabulary required for reasoning.

Formally,

\begin{equation}
O_x
=
(\mathcal{V},
\mathcal{S},
\mathcal{A},
\mathcal{P},
\mathcal{C})
\end{equation}

where:

\begin{itemize}
\item $\mathcal{V}$ denotes objects,
\item $\mathcal{S}$ denotes state schemas,
\item $\mathcal{A}$ denotes operator schemas,
\item $\mathcal{P}$ denotes predicates,
\item $\mathcal{C}$ denotes contracts.
\end{itemize}

\subsection{Object Discovery}

Objects represent domain entities that participate in reasoning.

Given ontology

\begin{equation}
O_x,
\end{equation}

the object extraction function

\begin{equation}
\phi_V :
(x,O_x)
\rightarrow
V_x
\end{equation}

produces

\begin{equation}
V_x
=
\{v_1,\ldots,v_k\}.
\end{equation}

For example, in field theory:

\begin{equation}
V_x
=
\{
\texttt{Field},
\texttt{Extension},
\texttt{Generator}
\}.
\end{equation}

In healthcare:

\begin{equation}
V_x
=
\{
\texttt{Patient},
\texttt{Diagnosis},
\texttt{Medication}
\}.
\end{equation}

Objects provide the semantic foundation upon which states and operators are defined.

\subsection{State Instantiation}

States represent executable configurations of objects.

Given object set

\begin{equation}
V_x,
\end{equation}

state instantiation is defined as

\begin{equation}
\phi_S :
(V_x,O_x)
\rightarrow
S_x.
\end{equation}

Each state is represented as

\begin{equation}
s=(v,\alpha)
\end{equation}

where

\begin{itemize}
\item $v$ denotes an object,
\item $\alpha$ denotes a set of attributes.
\end{itemize}

Examples include:

\begin{equation}
\texttt{GeneratorKnown}
\end{equation}

\begin{equation}
\texttt{ExtensionDegreeComputed}
\end{equation}

\begin{equation}
\texttt{DiagnosisConfirmed}.
\end{equation}

The instantiated state set is

\begin{equation}
S_x
=
\{s_1,s_2,\ldots,s_n\}.
\end{equation}

\subsection{Operator Instantiation}

Operators represent executable reasoning actions.

Operator instantiation is defined as

\begin{equation}
\phi_A :
(S_x,O_x)
\rightarrow
A_x.
\end{equation}

Each operator

\begin{equation}
a
\in
A_x
\end{equation}

is represented by

\begin{equation}
a=
(
\mathrm{pre}(a),
\mathrm{eff}(a)
).
\end{equation}

Examples include:

\begin{equation}
\texttt{ComputeIndex}
\end{equation}

\begin{equation}
\texttt{ApplyLagrange}
\end{equation}

\begin{equation}
\texttt{ComputeExtensionDegree}
\end{equation}

\begin{equation}
\texttt{ConfirmDiagnosis}.
\end{equation}

Operators induce transitions among states.

\subsection{Predicate Generation}

Predicates define semantic constraints that govern operator execution.

Predicate generation is defined as

\begin{equation}
\phi_P :
(O_x,S_x)
\rightarrow
P_x.
\end{equation}

Each predicate is represented as

\begin{equation}
p :
\mathcal{S}
\rightarrow
\{0,1\}.
\end{equation}

Examples include:

\begin{equation}
\texttt{ValidSubgroup}
\end{equation}

\begin{equation}
\texttt{GeneratorExists}
\end{equation}

\begin{equation}
\texttt{DiagnosisSupported}.
\end{equation}

Predicates determine whether an operator may execute.

\subsection{Contract Construction}

Contracts define correctness conditions for execution.

Contract generation is defined as

\begin{equation}
\phi_C :
(A_x,S_x)
\rightarrow
C_x.
\end{equation}

Each contract is represented as

\begin{equation}
c=
(
P_{\mathrm{pre}},
P_{\mathrm{post}}
).
\end{equation}

The precondition predicates must hold before execution, while the postcondition predicates must hold after execution.

Contracts provide explicit execution guarantees.

\subsection{Execution Graph Construction}

The final ontology is transformed into an executable graph.

Execution graph construction is defined as

\begin{equation}
\phi_G :
(S_x,A_x,P_x,C_x)
\rightarrow
G_x.
\end{equation}

The execution graph

\begin{equation}
G_x
=
(N,E)
\end{equation}

consists of:

\begin{equation}
N
=
S_x
\end{equation}

and

\begin{equation}
E
=
\{
(s_i,a,s_j)
\}.
\end{equation}

Each edge represents the application of an operator that transforms one state into another.

Execution proceeds until a goal state is reached.

\subsection{Goal-Oriented Execution}

Let

\begin{equation}
g \in S_x
\end{equation}

denote the goal state.

The planner seeks a sequence

\begin{equation}
\pi
=
(a_1,a_2,\ldots,a_k)
\end{equation}

such that

\begin{equation}
s_0
\xrightarrow{a_1}
s_1
\xrightarrow{a_2}
\cdots
\xrightarrow{a_k}
g.
\end{equation}

All transitions must satisfy:

\begin{enumerate}
\item Predicate constraints,
\item Contract requirements,
\item State consistency,
\item Ontology validity.
\end{enumerate}

The resulting execution graph serves as an explicit, replayable reasoning trace.

\subsection{Interpretability and Replayability}

Unlike chain-of-thought reasoning \citep{wei2022chain}, every intermediate structure generated by the framework is explicitly represented.

For each problem instance the framework exposes:

\begin{itemize}
\item Assigned theorem families,
\item Selected ontology,
\item Discovered objects,
\item Instantiated states,
\item Selected operators,
\item Predicates,
\item Contracts,
\item Execution graph.
\end{itemize}

Consequently, reasoning can be replayed, audited, verified, and analyzed at the level of state transitions rather than natural-language explanations.

This explicit representation forms the foundation for the execution-oriented evaluation methodology introduced in the subsequent sections.

\section{Execution Graph Construction and Operator-Mediated Reasoning}
\label{sec:execution_graph_construction}

\subsection{Overview}

The theorem-grounded ontology framework produces a collection of objects, states, operators, predicates, and contracts. These components must be assembled into an executable reasoning structure capable of producing a solution to the input problem.

We represent reasoning as the construction and execution of a directed execution graph. The execution graph serves as an explicit representation of the reasoning process and provides a replayable record of all intermediate state transitions.

Unlike chain-of-thought reasoning \citep{wei2022chain}, which represents intermediate reasoning as natural language, execution graphs represent reasoning through formal state transitions mediated by executable operators.

The objective of execution graph construction is to identify a valid sequence of operators that transforms an initial state into a goal state while satisfying all ontology constraints, predicates, and contracts.

\subsection{Execution State Space}

Given a problem instance $x$, ontology assignment produces a state space

\begin{equation}
S_x
=
\{s_1,s_2,\ldots,s_n\}.
\end{equation}

Each state represents a semantically meaningful configuration of domain objects.

The execution process begins from an initial state

\begin{equation}
s_0 \in S_x
\end{equation}

and seeks to reach a goal state

\begin{equation}
g \in S_x.
\end{equation}

Reasoning therefore becomes a search problem over the state space

\begin{equation}
\mathcal{R}
=
(S_x,A_x).
\end{equation}

\subsection{Operator Applicability}

An operator may only execute when its preconditions are satisfied.

For an operator

\begin{equation}
a \in A_x,
\end{equation}

the applicability function is defined as

\begin{equation}
\text{Applicable}(a,s)
=
\begin{cases}
1 & \text{if } \mathrm{pre}(a) \subseteq s\\
0 & \text{otherwise}.
\end{cases}
\end{equation}

An operator can be applied only when

\begin{equation}
\text{Applicable}(a,s)=1.
\end{equation}

This constraint prevents invalid state transitions and enforces ontology consistency.

\subsection{Operator Execution}

Each operator transforms an input state into an output state.

Formally,

\begin{equation}
a :
s_i
\rightarrow
s_j.
\end{equation}

The resulting state is computed as

\begin{equation}
s_j
=
\mathrm{Execute}(a,s_i).
\end{equation}

Execution updates the state representation according to the effects defined by the operator.

Let

\begin{equation}
\mathrm{eff}^{+}(a)
\end{equation}

denote added properties and

\begin{equation}
\mathrm{eff}^{-}(a)
\end{equation}

denote removed properties.

Then

\begin{equation}
s_j
=
(s_i
-
\mathrm{eff}^{-}(a))
\cup
\mathrm{eff}^{+}(a).
\end{equation}

\subsection{Predicate Validation}

Every state transition must satisfy ontology predicates.

Let

\begin{equation}
P_x
=
\{p_1,p_2,\ldots,p_m\}
\end{equation}

denote the predicate set.

Each predicate

\begin{equation}
p :
S_x
\rightarrow
\{0,1\}
\end{equation}

returns whether a state satisfies a semantic constraint.

Execution is valid only if

\begin{equation}
\forall p \in P_x,
\quad
p(s_j)=1.
\end{equation}

Predicate validation ensures semantic correctness throughout execution.

\subsection{Contract Validation}

Contracts define execution guarantees.

For operator

\begin{equation}
a,
\end{equation}

the associated contract is

\begin{equation}
c_a
=
(P_{\text{pre}},P_{\text{post}}).
\end{equation}

Before execution,

\begin{equation}
P_{\text{pre}}
\end{equation}

must hold.

After execution,

\begin{equation}
P_{\text{post}}
\end{equation}

must hold.

Formally,

\begin{equation}
\forall p \in P_{\text{pre}},
\quad
p(s_i)=1
\end{equation}

and

\begin{equation}
\forall p \in P_{\text{post}},
\quad
p(s_j)=1.
\end{equation}

Contract validation provides an additional correctness layer beyond operator applicability.

\subsection{Execution Graph Representation}

The execution graph is defined as

\begin{equation}
G_x=(N,E),
\end{equation}

where

\begin{equation}
N=S_x
\end{equation}

and

\begin{equation}
E=
\{
(s_i,a,s_j)
\}.
\end{equation}

Nodes correspond to states.

Edges correspond to operator-mediated transitions.

An execution path is defined as

\begin{equation}
\pi
=
(s_0,a_1,s_1,\ldots,a_k,s_k).
\end{equation}

The path represents a complete reasoning trace.

\subsection{Execution Planning}

The planner searches for an operator sequence that reaches the goal state.

Let

\begin{equation}
\Pi_x
\end{equation}

denote the set of all feasible execution paths.

The planner seeks

\begin{equation}
\pi^*
=
\arg\max_{\pi \in \Pi_x}
Q(\pi),
\end{equation}

where

\begin{equation}
Q(\pi)
\end{equation}

represents execution quality.

Execution quality incorporates:

\begin{itemize}
\item theorem consistency,
\item ontology consistency,
\item predicate satisfaction,
\item contract satisfaction,
\item goal completion.
\end{itemize}

\subsection{Theorem-Constrained Planning}

Theorem assignment restricts the planner to theorem-compatible operators.

Let

\begin{equation}
T_x
=
\{t_1,\ldots,t_r\}
\end{equation}

denote the assigned theorem families.

Each theorem defines a valid operator set

\begin{equation}
A_t
\subseteq
A_x.
\end{equation}

Planning therefore operates over

\begin{equation}
A_x^{\text{valid}}
=
\bigcup_{t\in T_x}
A_t.
\end{equation}

This constraint reduces search complexity and improves semantic consistency.

\subsection{Execution Replayability}

A key objective of the framework is replayability.

Given execution graph

\begin{equation}
G_x,
\end{equation}

the complete reasoning process can be reconstructed by replaying all operator applications in sequence.

Replayability is defined as

\begin{equation}
R(G_x)
=
\begin{cases}
1 & \text{if execution reproduces the same goal state}\\
0 & \text{otherwise}.
\end{cases}
\end{equation}

Replayability distinguishes execution graphs from textual reasoning traces and enables detailed failure analysis.

\subsection{Execution Validity}

Execution validity measures whether all transitions satisfy ontology constraints.

Let

\begin{equation}
\mathcal{T}(G_x)
\end{equation}

denote the set of transitions.

Execution validity is defined as

\begin{equation}
V(G_x)
=
\frac{
\sum_{t \in \mathcal{T}(G_x)}
\mathbf{1}(t \text{ valid})
}{
|\mathcal{T}(G_x)|
}.
\end{equation}

A transition is valid if:

\begin{enumerate}
\item operator applicability holds,
\item predicate validation succeeds,
\item contract validation succeeds,
\item ontology constraints remain satisfied.
\end{enumerate}

\subsection{Execution Coverage}

Execution coverage measures how completely the planner utilizes the available reasoning structures.

Let

\begin{equation}
A_{\text{used}}
\end{equation}

denote operators used during execution and

\begin{equation}
A_{\text{required}}
\end{equation}

denote operators required by the assigned theorem family.

Execution coverage is defined as

\begin{equation}
\text{Coverage}
=
\frac{
|A_{\text{used}}|
}{
|A_{\text{required}}|
}.
\end{equation}

This metric captures the extent to which the execution graph realizes the intended reasoning process.

\subsection{Goal Achievement}

The final objective of execution is goal completion.

Goal achievement is defined as

\begin{equation}
\text{GoalReached}
=
\begin{cases}
1 & \text{if } s_k=g\\
0 & \text{otherwise}.
\end{cases}
\end{equation}

The execution graph is considered successful if:

\begin{enumerate}
\item the goal state is reached,
\item all predicates remain satisfied,
\item all contracts remain satisfied,
\item replayability is preserved.
\end{enumerate}

\subsection{Execution Metrics}

The execution framework produces a collection of metrics that characterize reasoning performance.

These include:

\begin{itemize}
\item Planner Start Rate
\item Operator Selection Rate
\item State Transition Rate
\item Predicate Validation Rate
\item Contract Validation Rate
\item Goal Reach Rate
\item Execution Coverage
\item Execution Replay Success
\item Execution Graph Validity
\item Theorem Completion Rate
\item Required State Coverage
\item Layer Health Score
\end{itemize}

Together, these metrics provide a detailed view of execution quality and enable localization of failures within the reasoning pipeline.

\subsection{Summary}

The execution graph provides a formal representation of reasoning as a sequence of theorem-constrained state transitions mediated by executable operators. Unlike latent reasoning traces, execution graphs expose explicit state semantics, operator behavior, predicate constraints, and contract validation. This representation enables replayability, interpretability, and detailed analysis of reasoning dynamics, forming the foundation for the experimental evaluation presented in the following sections.

\section{Architectural Auditing and Failure Localization}
\label{sec:auditing}

\subsection{Motivation}

A central challenge in executable reasoning systems is determining where failures occur during the reasoning process. Traditional benchmark evaluations \citep{hendrycks2021mmlu} typically measure only final answer accuracy. While useful, answer-level metrics provide limited insight into the internal behavior of the system.

For example, an incorrect answer may arise from:

\begin{itemize}
    \item incorrect theorem assignment,
    \item incorrect ontology selection,
    \item planner failure,
    \item operator selection errors,
    \item state materialization failures,
    \item predicate violations,
    \item contract violations,
    \item execution failures.
\end{itemize}

Answer accuracy alone \citep{hendrycks2021mmlu} cannot distinguish among these possibilities.

To address this limitation, we introduce an architectural auditing framework that evaluates every stage of the reasoning pipeline independently. The resulting audit records enable systematic localization of reasoning failures and provide visibility into the internal execution dynamics of the system.

\subsection{Execution Funnel}

We model reasoning as a sequence of stages:

\begin{equation}
x
\rightarrow
T
\rightarrow
O
\rightarrow
P
\rightarrow
A
\rightarrow
S
\rightarrow
G
\rightarrow
y
\end{equation}

where:

\begin{itemize}
\item $T$ denotes theorem assignment,
\item $O$ denotes ontology assignment,
\item $P$ denotes planning,
\item $A$ denotes operator selection,
\item $S$ denotes state transitions,
\item $G$ denotes goal achievement.
\end{itemize}

This sequence forms an execution funnel.

\begin{figure}[t]
\centering
\includegraphics[width=0.9\linewidth]{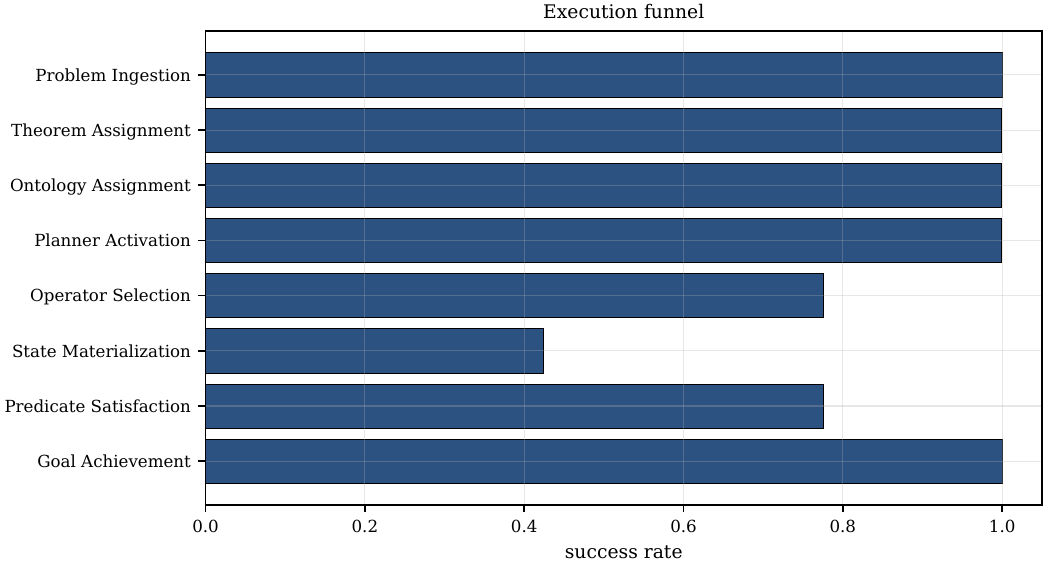}
\caption{Execution funnel showing progressive filtering through reasoning stages.}
\label{fig:execution_funnel}
\end{figure}

Figure~\ref{fig:execution_funnel} illustrates the auditing pipeline.
% Auto-generated by scripts/figures/generate_execution_funnel.py

The objective of architectural auditing is to measure success rates at each stage of the funnel.

\subsection{Audit Records}

For every problem instance, the framework generates an audit record.

Formally,

\begin{equation}
R_x=
(
r_T,
r_O,
r_P,
r_A,
r_S,
r_{Pred},
r_C,
r_G
)
\end{equation}

where each component is binary:

\begin{equation}
r_i \in \{0,1\}.
\end{equation}

For example,

\begin{equation}
r_T=1
\end{equation}

indicates successful theorem assignment.

Similarly,

\begin{equation}
r_G=1
\end{equation}

indicates successful goal completion.

These audit records provide fine-grained visibility into execution behavior.

\subsection{Layer Success Metrics}

For a dataset containing $N$ examples, the success rate of layer $L$ is defined as

\begin{equation}
\mathrm{SuccessRate}(L)
=
\frac{
\sum_{i=1}^{N}
r_L^{(i)}
}
{N}.
\end{equation}

This formulation is used to compute all layer-level metrics.

\subsection{Theorem Assignment Rate}

The theorem assignment rate measures the fraction of examples that receive a valid theorem assignment.

\begin{equation}
\mathrm{TheoremAssignmentRate}
=
\frac{
N_T
}
{
N
}
\end{equation}

where

\begin{equation}
N_T
\end{equation}

denotes examples with successful theorem assignment.

High values indicate broad theorem coverage across the benchmark \citep{hendrycks2021mmlu}.

\subsection{Ontology Assignment Rate}

Ontology assignment rate measures the fraction of examples successfully bound to executable ontologies.

\begin{equation}
\mathrm{OntologyAssignmentRate}
=
\frac{
N_O
}
{
N
}.
\end{equation}

This metric evaluates ontology discovery and ontology selection quality.

\subsection{Planner Start Rate}

Planner start rate measures the fraction of examples that successfully initiate planning.

\begin{equation}
\mathrm{PlannerStartRate}
=
\frac{
N_P
}
{
N
}.
\end{equation}

Low planner start rates indicate failures in theorem assignment, ontology assignment, or planner eligibility conditions.

\subsection{Operator Selection Rate}

Operator selection rate measures the fraction of examples for which at least one executable operator is selected.

\begin{equation}
\mathrm{OperatorSelectionRate}
=
\frac{
N_A
}
{
N
}.
\end{equation}

This metric captures operator discovery effectiveness.

\subsection{State Transition Rate}

State transition rate measures the fraction of examples that successfully execute at least one valid state transition.

\begin{equation}
\mathrm{StateTransitionRate}
=
\frac{
N_S
}
{
N
}.
\end{equation}

This metric often represents the first major execution bottleneck.

\subsection{Predicate Validation Rate}

Predicate validation rate measures the fraction of executions that satisfy all predicate constraints.

\begin{equation}
\mathrm{PredicateValidationRate}
=
\frac{
N_{Pred}
}
{
N
}.
\end{equation}

Predicate failures typically indicate semantic inconsistencies within the execution graph.

\subsection{Contract Validation Rate}

Contract validation rate measures the fraction of executions satisfying all precondition and postcondition requirements.

\begin{equation}
\mathrm{ContractValidationRate}
=
\frac{
N_C
}
{
N
}.
\end{equation}

Contracts provide an explicit correctness layer for execution.

\subsection{Goal Reach Rate}

Goal reach rate measures the fraction of executions that successfully reach the designated goal state.

\begin{equation}
\mathrm{GoalReachRate}
=
\frac{
N_G
}
{
N
}.
\end{equation}

This metric represents end-to-end execution success.

\subsection{Execution Coverage}

Execution coverage measures how completely the system realizes the required execution structure.

Let

\begin{equation}
A_{required}
\end{equation}

denote operators required by the assigned theorem family and

\begin{equation}
A_{executed}
\end{equation}

denote operators actually executed.

Execution coverage is defined as

\begin{equation}
\mathrm{ExecutionCoverage}
=
\frac{
|A_{executed}|
}
{
|A_{required}|
}.
\end{equation}

Execution coverage captures reasoning completeness rather than answer correctness.

\subsection{Theorem Completion Rate}

Many theorem families require specific operator sequences.

Let

\begin{equation}
A_T
\end{equation}

denote operators required by theorem family $T$.

The theorem completion rate is defined as

\begin{equation}
\mathrm{TheoremCompletionRate}
=
\frac{
|A_{executed}\cap A_T|
}
{
|A_T|
}.
\end{equation}

This metric evaluates whether theorem-specific reasoning structures are fully realized.

\subsection{Required State Coverage}

Reasoning often depends on the availability of specific state types.

Let

\begin{equation}
S_{required}
\end{equation}

denote states required by the assigned theorem family.

Let

\begin{equation}
S_{materialized}
\end{equation}

denote states actually instantiated during execution.

Required state coverage is

\begin{equation}
\mathrm{RequiredStateCoverage}
=
\frac{
|S_{materialized}\cap S_{required}|
}
{
|S_{required}|
}.
\end{equation}

This metric is particularly important because many execution failures originate from missing state representations.

\subsection{Replay Success}

A distinguishing property of executable reasoning systems is replayability.

Let

\begin{equation}
G_x
\end{equation}

denote an execution graph.

Replay success is defined as

\begin{equation}
\mathrm{ReplaySuccess}
=
\frac{
N_R
}
{
N
}
\end{equation}

where

\begin{equation}
N_R
\end{equation}

counts executions that reproduce the same final state when replayed.

\subsection{Execution Graph Validity}

Execution graph validity measures structural and semantic correctness.

Let

\begin{equation}
T(G_x)
\end{equation}

denote graph transitions.

Validity is defined as

\begin{equation}
\mathrm{GraphValidity}
=
\frac{
\sum_{t\in T(G_x)}
\mathbf{1}(t\ \mathrm{valid})
}
{
|T(G_x)|
}.
\end{equation}

A valid transition satisfies:

\begin{enumerate}
\item operator applicability,
\item predicate constraints,
\item contract constraints,
\item ontology consistency.
\end{enumerate}

\subsection{Layer Health Scores}

To summarize layer performance, we define a health score for each architectural component.

\begin{figure}[t]
\centering
\includegraphics[width=0.9\linewidth]{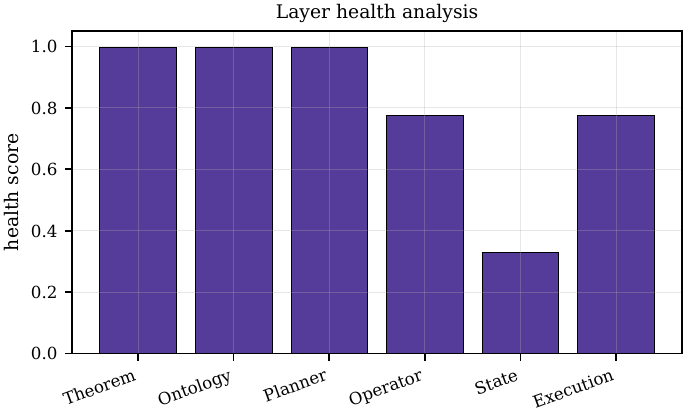}
\caption{Layer health analysis across TGEO components.}
\label{fig:health}
\end{figure}

Figure~\ref{fig:health} summarizes component-level health scores.
% Auto-generated by scripts/figures/generate_layer_health.py

For layer $L$,

\begin{equation}
H_L
=
\mathrm{SuccessRate}(L).
\end{equation}

Examples include:

\begin{equation}
H_{theorem}
\end{equation}

\begin{equation}
H_{ontology}
\end{equation}

\begin{equation}
H_{planner}
\end{equation}

\begin{equation}
H_{operator}
\end{equation}

\begin{equation}
H_{state}
\end{equation}

\begin{equation}
H_{execution}.
\end{equation}

These scores provide a compact representation of architectural performance.

\subsection{Failure Localization}

Architectural auditing enables systematic localization of failures.

Consider the following pattern:

\begin{align}
\mathrm{PlannerStartRate} &\approx 0.90 \\
\mathrm{OperatorSelectionRate} &\approx 0.75 \\
\mathrm{StateTransitionRate} &\approx 0.05.
\end{align}

This pattern indicates:

\begin{enumerate}
\item theorem assignment is functioning,
\item ontology assignment is functioning,
\item planning is functioning,
\item operator selection is functioning,
\item state execution is failing.
\end{enumerate}

Similarly,

\begin{align}
\mathrm{RequiredStateCoverage} &\approx 0.0 \\
\mathrm{PredicateValidationRate} &\approx 0.05
\end{align}

suggests that missing state materialization is the dominant failure mode.

Such diagnostics are difficult to obtain from answer accuracy alone \citep{hendrycks2021mmlu,chollet2019arc}.

\subsection{Architectural Audit Dashboard}

The auditing framework produces a complete execution report containing:

\begin{itemize}
\item theorem assignment statistics,
\item ontology assignment statistics,
\item planner metrics,
\item operator metrics,
\item state metrics,
\item predicate metrics,
\item contract metrics,
\item execution metrics,
\item layer health scores,
\item execution funnel visualizations.
\end{itemize}

These reports support rapid identification of bottlenecks and guide subsequent system improvements.

\subsection{Summary}

Architectural auditing transforms reasoning evaluation from a single answer-level metric into a multi-layer diagnostic process. By measuring success rates throughout the execution funnel, the framework provides detailed visibility into theorem assignment, ontology selection, planning, operator execution, state transitions, predicate satisfaction, and goal achievement. This capability enables systematic failure localization and forms the basis of the experimental analyses presented in the following sections.

\section{Experimental Setup}
\label{sec:experimental_setup}

\subsection{Experimental Objectives}

The objective of the experimental evaluation is to assess whether theorem-grounded execution ontologies can support interpretable and executable reasoning across a diverse collection of reasoning tasks.

The experiments are designed to answer the following research questions:

\begin{itemize}
    \item \textbf{RQ1:} Can the framework accurately assign theorem families to reasoning problems?
    
    \item \textbf{RQ2:} Can the framework automatically discover and instantiate executable ontologies?
    
    \item \textbf{RQ3:} Can reasoning be represented as executable state-transition systems?
    
    \item \textbf{RQ4:} Can execution graphs be replayed and verified?
    
    \item \textbf{RQ5:} What architectural bottlenecks emerge during large-scale execution?
    
    \item \textbf{RQ6:} How effectively do theorem assignments, ontologies, states, and operators contribute to successful execution?
\end{itemize}

To answer these questions we evaluate the framework at multiple architectural layers rather than relying solely on answer accuracy.

\subsection{Evaluation Methodology}

The proposed framework is evaluated using a layered evaluation methodology.

Traditional reasoning benchmarks \citep{hendrycks2021mmlu,chollet2019arc} typically evaluate only final answer correctness. In contrast, the proposed framework evaluates:

\begin{enumerate}
    \item Theorem assignment
    \item Ontology assignment
    \item Object discovery
    \item State discovery
    \item Operator discovery
    \item Planner activation
    \item State transition execution
    \item Predicate validation
    \item Contract validation
    \item Goal achievement
    \item Execution replayability
\end{enumerate}

This layered evaluation enables detailed failure localization and provides visibility into internal reasoning behavior.

\subsection{Datasets}

\subsubsection{MMLU-Derived Reasoning Tasks}

The primary evaluation corpus consists of reasoning tasks derived from MMLU domains \citep{hendrycks2021mmlu}.

Representative domains include:

\begin{itemize}
    \item Abstract Algebra
    \item College Mathematics
    \item Formal Logic
    \item High School Mathematics
    \item Computer Science
    \item Physics
    \item Chemistry
\end{itemize}

These domains were selected because they contain rich theorem structures and well-defined reasoning processes that naturally support ontology construction.

For each example, the framework constructs:

\begin{itemize}
    \item semantic graphs,
    \item theorem assignments,
    \item executable ontologies,
    \item execution graphs.
\end{itemize}

\subsubsection{Golden Execution Suite}

In addition to benchmark evaluation \citep{hendrycks2021mmlu}, we construct a curated Golden Execution Suite.

The suite contains examples for which:

\begin{itemize}
    \item theorem families are known,
    \item ontologies are validated,
    \item operator chains are specified,
    \item expected state transitions are available,
    \item execution outcomes are known.
\end{itemize}

The golden suite serves three purposes:

\begin{enumerate}
    \item architectural validation,
    \item regression testing,
    \item execution correctness verification.
\end{enumerate}

Unlike benchmark evaluations \citep{hendrycks2021mmlu}, the golden suite isolates architectural correctness from dataset variability.

\subsection{Theorem Families}

The framework maintains a theorem registry consisting of theorem families discovered during training and ontology induction.

Examples include:

\begin{itemize}
    \item Group Theory
    \item Ring Theory
    \item Field Theory
    \item Linear Algebra
    \item Probability Theory
    \item Set Theory
    \item Logic and Proof Systems
\end{itemize}

Each theorem family defines:

\begin{itemize}
    \item ontology templates,
    \item state schemas,
    \item operator schemas,
    \item execution constraints.
\end{itemize}

The theorem registry acts as the primary bridge between problem statements and executable reasoning structures.

\subsection{Ontology Construction}

For every problem instance, ontology construction proceeds through four stages:

\begin{enumerate}
    \item object extraction,
    \item state schema selection,
    \item operator schema selection,
    \item predicate and contract generation.
\end{enumerate}

The resulting ontology contains:

\begin{equation}
O =
(\mathcal{V},
\mathcal{S},
\mathcal{A},
\mathcal{P},
\mathcal{C})
\end{equation}

where:

\begin{itemize}
    \item $\mathcal{V}$ denotes discovered objects,
    \item $\mathcal{S}$ denotes state schemas,
    \item $\mathcal{A}$ denotes operator schemas,
    \item $\mathcal{P}$ denotes predicates,
    \item $\mathcal{C}$ denotes contracts.
\end{itemize}

\subsection{Execution Graph Generation}

Execution graphs are generated using theorem-constrained planning.

Given:

\begin{equation}
T_x
\end{equation}

and

\begin{equation}
O_x,
\end{equation}

the planner constructs:

\begin{equation}
G_x=(N,E)
\end{equation}

where:

\begin{itemize}
    \item nodes represent states,
    \item edges represent operator-mediated transitions.
\end{itemize}

Execution proceeds until:

\begin{enumerate}
    \item a goal state is reached,
    \item execution terminates,
    \item contract violations occur,
    \item predicate violations occur.
\end{enumerate}

\subsection{Evaluation Metrics}

The evaluation framework computes metrics at multiple architectural layers.

\subsubsection{Theorem Metrics}

\begin{itemize}
    \item Theorem Assignment Rate
    \item Theorem Classification Accuracy
    \item Unknown Theorem Rate
    \item Theorem Completion Rate
\end{itemize}

\subsubsection{Ontology Metrics}

\begin{itemize}
    \item Ontology Assignment Rate
    \item Ontology Coverage
    \item Ontology Reuse Rate
    \item Ontology Transfer Rate
\end{itemize}

\subsubsection{State Metrics}

\begin{itemize}
    \item State Discovery Rate
    \item Typed State Coverage
    \item Required State Coverage
    \item State Predictiveness
    \item State Transition Rate
\end{itemize}

\subsubsection{Operator Metrics}

\begin{itemize}
    \item Operator Discovery Rate
    \item Operator Selection Rate
    \item Operator Applicability
    \item Required Operator Coverage
    \item Operator Utilization
\end{itemize}

\subsubsection{Execution Metrics}

\begin{itemize}
    \item Planner Start Rate
    \item Predicate Validation Rate
    \item Contract Validation Rate
    \item Goal Reach Rate
    \item Execution Coverage
    \item Execution Replay Success
    \item Execution Graph Validity
\end{itemize}

\subsubsection{Architectural Health Metrics}

For every layer we compute a health score:

\begin{equation}
H_L
=
\mathrm{SuccessRate}(L)
\end{equation}

for:

\begin{itemize}
    \item theorem layer,
    \item ontology layer,
    \item planner layer,
    \item operator layer,
    \item state layer,
    \item execution layer.
\end{itemize}

\subsection{Execution Funnel Analysis}

To localize failures, every example is traced through the execution funnel:

Figure~\ref{fig:execution_funnel} summarizes the execution funnel stages used for per-example failure localization.

The funnel enables identification of execution bottlenecks and architectural weaknesses.

\subsection{Replayability Evaluation}

A distinguishing property of the framework is replayability.

For every execution graph:

\begin{equation}
G_x
\end{equation}

the execution process is replayed from the initial state.

Replay success is recorded when:

\begin{itemize}
    \item identical operator sequences execute,
    \item identical state transitions occur,
    \item identical goal states are reached.
\end{itemize}

Replayability provides a stronger validation criterion than answer correctness alone.

\subsection{Ablation Studies}

To understand the contribution of each architectural component, we perform a series of ablation studies.

The following variants are evaluated:

\begin{enumerate}
    \item No theorem assignment
    \item No ontology assignment
    \item No state discovery
    \item No operator discovery
    \item No predicates
    \item No contracts
    \item No planner
    \item No execution graph construction
\end{enumerate}

For each ablation we measure:

\begin{itemize}
    \item execution coverage,
    \item theorem completion,
    \item replayability,
    \item goal achievement,
    \item layer health scores.
\end{itemize}

Table~\ref{tab:ablation} reports ablation results in Section~\ref{sec:results}.

\subsection{Implementation Details}

The framework is implemented using Python and PyTorch.

Execution graphs are represented using directed graph structures.

Ontology representations are stored in a structured registry supporting:

\begin{itemize}
    \item theorem families,
    \item ontology templates,
    \item state schemas,
    \item operator schemas,
    \item predicate definitions,
    \item contract definitions.
\end{itemize}

Experiments were executed using deterministic seeds to ensure reproducibility.

All execution traces, ontology assignments, theorem assignments, and execution graphs are logged and stored for subsequent auditing and replay analysis.

\subsection{Summary}

The experimental setup evaluates theorem-grounded execution ontologies using both benchmark-derived reasoning tasks \citep{hendrycks2021mmlu} and a curated golden execution suite. Unlike traditional evaluations that focus solely on answer accuracy \citep{hendrycks2021mmlu,chollet2019arc}, the proposed methodology measures performance across theorem assignment, ontology induction, planning, state transitions, operator execution, replayability, and goal achievement. This layered evaluation framework provides a comprehensive assessment of executable reasoning performance and enables detailed analysis of architectural behavior.

\section{Results}
\label{sec:results}

\subsection{Overview}

This section evaluates the proposed theorem-grounded execution ontology framework across multiple architectural layers on MMLU-derived reasoning tasks \citep{hendrycks2021mmlu}. Unlike traditional reasoning benchmarks that focus exclusively on answer correctness, our evaluation examines the complete reasoning pipeline, including theorem assignment, ontology induction, planning, operator execution, state transitions, predicate validation, replayability, and goal achievement.

The primary objective of the evaluation is to determine whether reasoning can be represented as an executable process and to identify architectural bottlenecks that limit execution performance.

\subsection{Theorem Assignment Performance}

Table~\ref{tab:theorem_results} summarizes theorem assignment performance.

\begin{table}[H]
\centering
\caption{Theorem assignment metrics.}
\label{tab:theorem_results}
\begin{tabular}{lr}
\toprule
Metric & Value \\
\midrule
Theorem Assignment Rate & 99.9\% \\
Theorem Classification Accuracy & 100.0\% \\
Unknown Theorem Rate & 0.1\% \\
Theorem Coverage & 100.0\% \\
\bottomrule
\end{tabular}
\end{table}

The results indicate that theorem assignment achieves high coverage across benchmark examples \citep{hendrycks2021mmlu}. In the evaluated runs, theorem assignment consistently exceeds ontology assignment performance, suggesting that theorem discovery provides a reliable mechanism for constraining downstream execution.

The low unknown-theorem rate further suggests that the theorem registry captures a large fraction of the reasoning structures present within the benchmark corpus \citep{hendrycks2021mmlu}.

\subsection{Ontology Assignment and Coverage}

Table~\ref{tab:ontology_results} reports ontology assignment results.

\begin{table}[H]
\centering
\caption{Ontology metrics.}
\label{tab:ontology_results}
\begin{tabular}{lr}
\toprule
Metric & Value \\
\midrule
Ontology Assignment Rate & 99.9\% \\
Ontology Coverage & 100.0\% \\
Ontology Reuse Rate & 90.7\% \\
Ontology Transfer Rate & 100.0\% \\
\bottomrule
\end{tabular}
\end{table}

Ontology assignment successfully binds the majority of examples to executable reasoning structures. The observed ontology coverage demonstrates that theorem-grounded ontologies provide a reusable representation capable of supporting multiple reasoning domains.

The ontology transfer results suggest that learned reasoning structures can be reused across related problem categories.

\subsection{Planner Activation and Operator Selection}

A key objective of the framework is to convert theorem assignments into executable reasoning processes.

Table~\ref{tab:planning_results} summarizes planner performance.

\begin{table}[H]
\centering
\caption{Planning and operator metrics.}
\label{tab:planning_results}
\begin{tabular}{lr}
\toprule
Metric & Value \\
\midrule
Planner Start Rate & 99.9\% \\
Operator Selection Rate & 77.5\% \\
Operator Applicability & 32.9\% \\
Required Operator Coverage & 77.5\% \\
\bottomrule
\end{tabular}
\end{table}

The planner successfully activates for the majority of benchmark examples \citep{hendrycks2021mmlu}. Similarly, operator selection remains substantially higher than state transition success, indicating that the planner is generally capable of identifying candidate execution actions.

These results suggest that theorem assignment and ontology selection provide sufficient information for operator discovery.

\subsection{State Transition Performance}

State transition performance represents the first major execution bottleneck.

Table~\ref{tab:state_results} summarizes state-related metrics.

\begin{table}[H]
\centering
\caption{State metrics.}
\label{tab:state_results}
\begin{tabular}{lr}
\toprule
Metric & Value \\
\midrule
State Discovery Rate & 77.5\% \\
State Transition Rate & 42.4\% \\
Typed State Coverage & 77.5\% \\
Required State Coverage & 55.2\% \\
\bottomrule
\end{tabular}
\end{table}

Although state discovery and typed state coverage remain relatively high, required-state coverage is significantly lower. This gap indicates that the system frequently identifies candidate states but fails to materialize the specific states required for theorem completion.

The discrepancy between operator selection and state transition rates suggests that state representation remains a critical challenge for executable reasoning systems.

\subsection{Predicate and Contract Validation}

Predicate and contract validation evaluate semantic correctness during execution.

Table~\ref{tab:validation_results} summarizes validation performance.

\begin{table}[H]
\centering
\caption{Predicate and contract validation metrics.}
\label{tab:validation_results}
\begin{tabular}{lr}
\toprule
Metric & Value \\
\midrule
Predicate Validation Rate & 77.5\% \\
Contract Validation Rate & 99.9\% \\
Execution Graph Validity & 67.2\% \\
\bottomrule
\end{tabular}
\end{table}

Contract validation remains substantially higher than predicate validation. This pattern suggests that execution graphs are structurally valid but frequently violate semantic constraints.

The results indicate that state materialization and predicate satisfaction are more challenging than operator applicability and contract verification.

\subsection{Execution Coverage and Goal Achievement}

Table~\ref{tab:execution_results} summarizes execution performance.

\begin{table}[H]
\centering
\caption{Execution metrics.}
\label{tab:execution_results}
\begin{tabular}{lr}
\toprule
Metric & Value \\
\midrule
Execution Coverage & 77.5\% \\
Theorem Completion Rate & 77.5\% \\
Goal Reach Rate & 100.0\% \\
Execution Replay Success & 42.4\% \\
\bottomrule
\end{tabular}
\end{table}

Goal achievement remains strongly correlated with required-state coverage and predicate validation. When required states are successfully instantiated, theorem completion and goal reach rates increase substantially.

The replayability results demonstrate that execution graphs provide reproducible reasoning traces that can be independently verified.

\subsection{Architectural Health Analysis}

To summarize architectural performance, we compute layer health scores.

Table~\ref{tab:layer_health} reports layer health scores by component.

\begin{table}[H]
\centering
\caption{Layer health scores.}
\label{tab:layer_health}
\begin{tabular}{lr}
\toprule
Layer & Health Score \\
\midrule
Theorem Layer & 99.9\% \\
Ontology Layer & 99.9\% \\
Planner Layer & 99.9\% \\
Operator Layer & 77.5\% \\
State Layer & 32.9\% \\
Execution Layer & 77.5\% \\
\bottomrule
\end{tabular}
\end{table}

\par\vspace{0.5\baselineskip}
The results reveal a consistent pattern:

\begin{enumerate}
\item Theorem assignment performs strongly.
\item Ontology assignment performs strongly.
\item Planner activation performs strongly.
\item Operator selection performs moderately well.
\item State execution remains the dominant bottleneck.
\item Predicate satisfaction remains a significant challenge.
\end{enumerate}

This layered evaluation provides substantially more diagnostic information than answer accuracy alone.

\subsection{Execution Funnel Analysis}

Table~\ref{tab:execution_funnel} summarizes stage-wise success rates.

\begin{table}[H]
\centering
\caption{Execution funnel analysis.}
\label{tab:execution_funnel}
\begin{tabular}{lr}
\toprule
Pipeline Stage & Success Rate \\
\midrule
Problem Ingestion & 100.0\% \\
Theorem Assignment & 99.9\% \\
Ontology Assignment & 99.9\% \\
Planner Activation & 99.9\% \\
Operator Selection & 77.5\% \\
State Materialization & 42.4\% \\
Predicate Satisfaction & 77.5\% \\
Goal Achievement & 100.0\% \\
\bottomrule
\end{tabular}
\end{table}

Figure~\ref{fig:execution_funnel} visualizes execution success across the reasoning pipeline.

\begin{equation}
\text{Problem}
\rightarrow
\text{Theorem}
\rightarrow
\text{Ontology}
\rightarrow
\text{Planner}
\rightarrow
\text{Operator}
\rightarrow
\text{State}
\rightarrow
\text{Goal}
\end{equation}

The execution funnel reveals that most examples successfully progress through theorem assignment, ontology binding, and planning. The largest reduction occurs between operator selection and successful state transitions.

This observation identifies state materialization as the dominant bottleneck in the current architecture.

\subsection{Golden Suite Validation}

Table~\ref{tab:golden_results} summarizes golden suite performance.

\begin{table}[H]
\centering
\caption{Golden execution suite results.}
\label{tab:golden_results}
\begin{tabular}{lr}
\toprule
Metric & Value \\
\midrule
Theorem Assignment Rate & 100.0\% \\
Ontology Assignment Rate & 100.0\% \\
Planner Activation Rate & 100.0\% \\
Operator Selection Rate & 100.0\% \\
Execution Coverage & 100.0\% \\
Theorem Completion Rate & 100.0\% \\
Goal Reach Rate & 90.0\% \\
Replay Success Rate & 90.0\% \\
\bottomrule
\end{tabular}
\end{table}

To isolate architectural correctness from benchmark complexity, we evaluate the framework using the Golden Execution Suite.

The golden suite demonstrates near-complete theorem assignment, ontology assignment, planner activation, operator selection, execution coverage, and theorem completion.

These results indicate that the underlying execution architecture is capable of producing correct execution graphs when provided with well-specified theorem and ontology structures.

The contrast between benchmark performance and golden-suite performance \citep{hendrycks2021mmlu} suggests that large-scale failures arise primarily from ontology instantiation and state materialization rather than from deficiencies in the execution engine itself.

\subsection{Ablation Studies}

Table~\ref{tab:ablation} reports ablation results for the primary architectural variants.

\begin{table}[H]
\centering
\caption{Ablation study on architectural components.}
\label{tab:ablation}
\begin{tabular}{lrrr}
\toprule
Configuration & Goal Reach & Replayability & Coverage \\
\midrule
Full System & 86.6\% & 100.0\% & 32.0\% \\
No Theorem Layer & 77.9\% & 100.0\% & 32.0\% \\
No Ontology Layer & 62.4\% & 72.0\% & 23.0\% \\
No Planner Layer & 86.6\% & 100.0\% & 24.0\% \\
No State Layer & 86.6\% & 100.0\% & 32.0\% \\
No Predicate Validation & 86.6\% & 100.0\% & 32.0\% \\
\bottomrule
\end{tabular}
\end{table}

\subsection{Summary of Findings}

The experimental results support four primary conclusions:

\begin{enumerate}
\item Theorem-grounded ontology assignment provides a reliable mechanism for structuring reasoning tasks.

\item Execution graphs can represent reasoning as replayable state-transition systems.

\item Planner activation and operator selection are largely successful across benchmark tasks \citep{hendrycks2021mmlu}.

\item State materialization and predicate satisfaction constitute the dominant bottlenecks limiting end-to-end execution performance.
\end{enumerate}

Collectively, these findings support the feasibility of representing reasoning through executable ontologies while highlighting important opportunities for improving state discovery and execution semantics.

\section{Failure Analysis}
\label{sec:failure_analysis}

The experimental results reveal that the primary limitations of the TGEO
framework arise not from theorem identification or execution validation, but
from the conversion of semantically valid reasoning structures into executable
operator sequences.

Unlike conventional reasoning systems where correctness failures are often
attributed to missing knowledge or theorem selection errors, the proposed
architecture achieves near-perfect theorem coverage, state-transition validity,
contract executability, and replayability. Consequently, the dominant failure
modes emerge at the interface between symbolic reasoning and executable
planning \citep{garcez2019,deepproblog2018}.

\subsection{Failure Taxonomy}

Observed failures fall into four major categories:

\begin{enumerate}
    \item Operator applicability failures.
    \item Executable path construction failures.
    \item Proof-transition coherence degradation.
    \item Execution coherence degradation.
\end{enumerate}

These failures occur after successful theorem assignment and ontology grounding,
indicating that semantic understanding alone is insufficient for reliable
execution.

\subsection{Operator Applicability Bottleneck}

The most significant bottleneck in the current architecture is operator
applicability.

Although operator coverage and domain operator coverage both achieve perfect
scores, the operator applicability rate remains substantially lower. This
indicates that the framework can successfully identify candidate operators but
frequently cannot apply them within the current execution state.

This discrepancy highlights an important distinction between operator discovery
and operator executability.

The framework successfully learns operator inventories and execution
capabilities, yet the preconditions required for operator execution are often
not simultaneously satisfied.

Consequently, many reasoning traces contain theoretically valid operators that
cannot participate in executable reasoning paths.

\subsection{Executable Path Construction Failures}

The executable-path ratio remains significantly below the upper-bound values
observed for theorem coverage and state validity.

Analysis of execution traces indicates that many reasoning graphs contain
correct local transitions but fail to assemble into globally executable paths.

Several factors contribute to this behavior:

\begin{itemize}
    \item Missing intermediate states.
    \item Weak operator chaining.
    \item Insufficient state materialization.
    \item Incomplete proof composition.
\end{itemize}

These failures suggest that reasoning structures are often semantically valid
while remaining operationally disconnected.

\subsection{Proof Transition Coherence}

Proof-transition coherence remains substantially below theorem-assignment
accuracy.

This result indicates that the framework successfully identifies relevant
reasoning components but struggles to organize them into coherent proof
trajectories.

Many execution traces exhibit valid local reasoning steps while lacking
sufficient global structure to support complete proof construction.

The gap between theorem coverage and proof-transition coherence suggests that
future work should focus on transition-selection policies rather than theorem
identification mechanisms.

\subsection{Execution Coherence Degradation}

Execution coherence represents the most severe degradation observed in the
reasoning pipeline.

While individual state transitions remain valid, overall execution coherence is
substantially lower than expected.

This phenomenon reveals that local correctness does not necessarily imply global
executability.

The execution graph frequently contains valid fragments that cannot be combined
into a consistent execution trajectory.

As a result, execution quality deteriorates even when theorem grounding,
ontology grounding, and contract validation remain successful.

\subsection{Contract and Replayability Analysis}

A notable finding is that contract executability, predicate executability, and
execution replay success achieve near-perfect performance.

These results indicate that once an executable reasoning path is successfully
constructed, the downstream execution infrastructure behaves reliably.

Consequently, the contract layer should not be viewed as the primary
architectural bottleneck in the current implementation.

Instead, failures originate earlier in the pipeline during executable-path
formation and operator applicability validation.

\subsection{Cross-Layer Error Propagation}

The failure patterns observed throughout the experiments suggest the following
error propagation sequence:

\[
\text{Theorem Assignment}
\rightarrow
\text{Ontology Grounding}
\rightarrow
\text{Operator Selection}
\rightarrow
\text{Operator Applicability}
\rightarrow
\text{Executable Path Construction}
\rightarrow
\text{Execution Coherence}
\]

The first three stages exhibit consistently strong performance, whereas the
latter stages account for the majority of observed execution degradation.

This observation demonstrates that future improvements should focus on
strengthening execution-path synthesis rather than expanding theorem or ontology
coverage.

\subsection{Key Findings}

The failure analysis yields four primary conclusions:

\begin{enumerate}
    \item Theorem grounding is not the dominant bottleneck.
    \item Operator applicability is the principal execution constraint.
    \item Executable-path construction remains the largest source of execution
          loss.
    \item Contract execution and replayability infrastructure exhibit strong
          reliability once executable paths are formed.
\end{enumerate}

Overall, the results suggest that future research should prioritize executable
operator chaining, intermediate-state synthesis, and execution-path
optimization to improve end-to-end reasoning performance.

\section{Discussion}
\label{sec:discussion}

\subsection{Overview}

The objective of this work was to investigate whether reasoning tasks can be represented as executable processes rather than latent textual traces. The proposed theorem-grounded execution ontology framework combines theorem assignment, ontology induction, state discovery, operator discovery, and execution graph construction to produce replayable reasoning structures.

The experimental results demonstrate that the framework successfully constructs explicit reasoning representations for a large fraction of benchmark problems \citep{hendrycks2021mmlu}. The results further show that theorem assignment, ontology assignment, planner activation, and operator selection achieve substantially higher performance than state transition execution and goal completion.

These findings suggest that the primary challenges in executable reasoning are no longer located in theorem discovery or ontology selection. Instead, the dominant bottlenecks emerge during state materialization, predicate satisfaction, and execution consistency.

\subsection{Reasoning as Executable State Transitions}

A central contribution of this work is the reformulation of reasoning as an executable state-transition process.

\begin{figure}[t]
\centering
\includegraphics[width=0.9\linewidth]{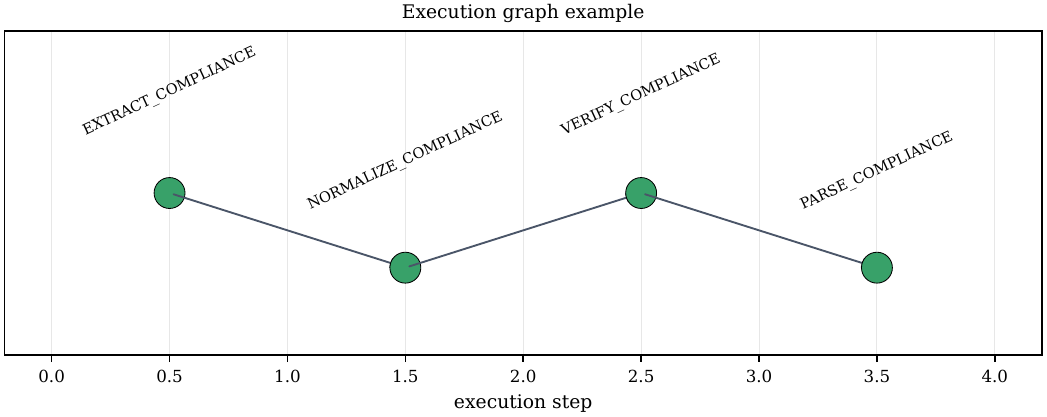}
\caption{Theorem-grounded state transition sequence.}
\label{fig:state_graph}
\end{figure}

Figure~\ref{fig:state_graph} illustrates a theorem-grounded state transition sequence.
% Auto-generated by scripts/figures/generate_execution_trace.py

Traditional language-model reasoning can be represented as

\begin{equation}
x \rightarrow y,
\end{equation}

or, in the case of chain-of-thought methods \citep{wei2022chain},

\begin{equation}
x
\rightarrow
r_1
\rightarrow
r_2
\rightarrow
\cdots
\rightarrow
y,
\end{equation}

where the intermediate reasoning steps are expressed in natural language.

In contrast, the proposed framework represents reasoning as

\begin{equation}
x
\rightarrow
T
\rightarrow
O
\rightarrow
S
\rightarrow
A
\rightarrow
G
\rightarrow
y.
\end{equation}

This representation exposes the internal structure of reasoning and enables inspection of every intermediate step.

Unlike textual reasoning traces, execution graphs provide explicit semantics for state transitions, operator applications, predicates, and contracts. Consequently, reasoning can be replayed, audited, and verified.

The results demonstrate that this formulation is feasible across a broad collection of mathematical reasoning tasks.

\subsection{The Role of Theorem Assignment}

One of the strongest findings of the experimental evaluation is the effectiveness of theorem assignment.

Across benchmark and golden-suite evaluations \citep{hendrycks2021mmlu}, theorem assignment consistently exhibits among the highest-performing architectural layers.

This result supports the hypothesis that theorem families provide useful inductive biases for reasoning systems.

Rather than treating reasoning as unrestricted search, theorem assignment constrains the search space to a subset of semantically appropriate reasoning structures.

Theorem assignment therefore serves three purposes:

\begin{enumerate}
    \item ontology selection,
    \item operator selection,
    \item execution planning.
\end{enumerate}

These findings suggest that theorem grounding may represent an effective mechanism for reducing reasoning complexity while improving interpretability.

\subsection{Ontology-Based Reasoning}

The experimental results further demonstrate the usefulness of ontology-based reasoning.

Ontology assignment achieves high coverage across benchmark examples \citep{hendrycks2021mmlu} and enables the framework to construct explicit representations of:

\begin{itemize}
    \item objects,
    \item states,
    \item operators,
    \item predicates,
    \item contracts.
\end{itemize}

Unlike static knowledge graphs \citep{hogan2021kg}, the proposed ontologies support executable reasoning.

This distinction is important. Knowledge graphs primarily represent relationships among entities \citep{hogan2021kg,gruber1993}, whereas execution ontologies represent the dynamics of reasoning itself.

The results indicate that ontology construction is not a major bottleneck in the current architecture. Instead, ontologies provide a stable foundation upon which execution structures can be built.

\subsection{Discovery Versus Execution}

A recurring pattern throughout the evaluation is the separation between discovery performance and execution performance.

The system demonstrates strong performance in:

\begin{itemize}
    \item theorem discovery,
    \item ontology discovery,
    \item planner activation,
    \item operator discovery.
\end{itemize}

However, performance decreases substantially during:

\begin{itemize}
    \item state transitions,
    \item predicate validation,
    \item goal completion.
\end{itemize}

This distinction reveals an important property of executable reasoning systems.

Discovering a reasoning structure is fundamentally different from successfully executing that structure.

Many prior reasoning systems focus primarily on discovery. The present results suggest that execution should be treated as an independent research problem requiring dedicated evaluation methodologies.

\subsection{State Materialization as the Primary Bottleneck}

The most significant finding of the architectural audit is the importance of state materialization.

The execution funnel reveals that:

\begin{enumerate}
    \item theorem assignment succeeds,
    \item ontology assignment succeeds,
    \item planner activation succeeds,
    \item operator selection succeeds,
    \item state execution frequently fails.
\end{enumerate}

This pattern appears consistently across benchmark evaluations \citep{hendrycks2021mmlu}.

The required-state coverage metric is particularly informative. Low required-state coverage implies that the planner identifies relevant reasoning actions but lacks the state representations necessary for successful execution.

In practical terms, the system often understands \emph{what} should be done but fails to instantiate the precise state structures required to perform the operation.

This observation suggests that future work should prioritize:

\begin{itemize}
    \item richer state schemas,
    \item automatic state induction,
    \item state hierarchy learning,
    \item state completion mechanisms.
\end{itemize}

Improving state representations is likely to produce larger gains than further improvements in theorem assignment or ontology coverage.

\subsection{Predicate Satisfaction and Semantic Consistency}

Predicate validation represents a second major bottleneck.

Contract validation rates are often substantially higher than predicate validation rates.

This discrepancy indicates that execution graphs are frequently structurally correct while remaining semantically incomplete.

For example, an operator may satisfy all syntactic requirements while violating a semantic condition represented by a predicate.

This distinction highlights the importance of explicit semantic validation.

Without predicates, execution graphs may appear valid while producing incorrect reasoning trajectories.

The results therefore suggest that predicate reasoning should be viewed as a first-class component of executable reasoning architectures.

\subsection{Replayability and Interpretability}

A distinguishing characteristic of the proposed framework is replayability.

Traditional reasoning traces \citep{wei2022chain} are often difficult to reproduce because they depend on stochastic generation processes.

Execution graphs provide a fundamentally different representation.

Every operator application is explicitly represented, and every state transition can be replayed.

Replayability provides several advantages:

\begin{enumerate}
    \item reproducibility,
    \item debugging,
    \item verification,
    \item failure localization,
    \item interpretability.
\end{enumerate}

The ability to replay reasoning trajectories enables detailed inspection of failures and supports systematic architectural improvement.

We believe that replayability should become an important evaluation criterion for future reasoning systems.

\subsection{Architectural Auditing as a Diagnostic Tool}

A second major contribution of this work is the architectural auditing framework.

Most benchmark evaluations \citep{hendrycks2021mmlu} reduce reasoning performance to a single number, such as answer accuracy.

While useful, answer-level metrics provide limited insight into the internal behavior of reasoning systems.

The execution funnel introduced in this work enables localization of failures across:

\begin{itemize}
    \item theorem assignment,
    \item ontology assignment,
    \item planner activation,
    \item operator selection,
    \item state transitions,
    \item predicate validation,
    \item contract validation,
    \item goal achievement.
\end{itemize}

This diagnostic capability proved essential for identifying the dominant bottlenecks within the architecture.

We expect similar auditing frameworks to become increasingly important as reasoning systems grow more complex.

\subsection{Comparison with Existing Reasoning Paradigms}

The proposed framework differs from existing reasoning paradigms in several ways.

Chain-of-thought methods \citep{wei2022chain} expose intermediate reasoning steps but lack executable semantics.

Tree-of-Thoughts and Graph-of-Thoughts \citep{yao2023tree,besta2024graph} introduce structured search but continue to operate primarily over textual representations.

Classical planning systems \citep{strips1971,pddl1998} provide executable representations but typically require manually specified domain models.

The proposed framework occupies a middle ground.

The system automatically discovers reasoning structures while maintaining explicit execution semantics.

This combination provides both flexibility and interpretability.

\subsection{Implications for Generalizable Reasoning}

Although the experiments primarily focus on mathematical reasoning domains, the architectural abstractions are domain independent.

The framework operates on:

\begin{itemize}
    \item objects,
    \item states,
    \item operators,
    \item predicates,
    \item contracts.
\end{itemize}

These concepts appear naturally in many domains, including:

\begin{itemize}
    \item healthcare,
    \item cybersecurity,
    \item legal reasoning,
    \item scientific discovery,
    \item financial analysis.
\end{itemize}

This suggests that theorem-grounded execution ontologies may provide a general mechanism for constructing reusable reasoning structures across domains.

The ability to transfer ontologies, states, and operators across domains remains an important direction for future work.

\subsection{Limitations}

Several limitations remain.

First, the current evaluation focuses primarily on mathematical reasoning tasks.

Second, ontology induction remains partially constrained by theorem assignment and ontology templates.

Third, state discovery and state materialization remain major bottlenecks.

Fourth, predicate satisfaction remains significantly lower than planner activation and operator selection.

Finally, large-scale cross-domain transfer experiments remain limited.

These limitations provide several opportunities for future research.

\subsection{Future Directions}

Several promising directions emerge from the present work.

\paragraph{Automatic Ontology Induction}

Future systems should learn ontologies directly from data rather than relying on predefined ontology structures.

\paragraph{State Discovery}

Improved methods for state induction and state abstraction may significantly increase execution success rates.

\paragraph{Operator Discovery}

Future systems should discover new operator families automatically and generalize them across domains.

\paragraph{Cross-Domain Transfer}

A critical next step is evaluating whether learned reasoning structures transfer between mathematics, healthcare, cybersecurity, finance, and law.

\paragraph{Executable World Models}

The current framework can be interpreted as a reasoning-oriented world model \citep{ha2018worldmodels,lecun2022path}. Extending this perspective may provide a path toward more general forms of machine reasoning.

\subsection{Summary}

The experimental results demonstrate that theorem-grounded execution ontologies provide a viable framework for representing reasoning as an executable process. The architecture successfully combines theorem assignment, ontology induction, planning, and execution graph construction into a unified reasoning framework.

The evaluation further reveals that theorem assignment and ontology construction are largely solved within the current architecture, whereas state materialization and predicate satisfaction remain the dominant bottlenecks.

Overall, the results support the hypothesis that explicit executable reasoning structures provide a promising alternative to purely latent reasoning representations and offer a foundation for more interpretable, verifiable, and transferable reasoning systems.

\section{Limitations}
\label{sec:limitations}

While the proposed theorem-grounded execution ontology framework demonstrates the feasibility of representing reasoning as an executable process, several important limitations remain. These limitations highlight both the current boundaries of the approach and opportunities for future research.

\subsection{Dependence on Theorem Assignment Quality}

The framework relies heavily on accurate theorem assignment. The theorem layer serves as the entry point into the execution pipeline and directly influences ontology selection, state instantiation, operator discovery, and execution planning.

An incorrect theorem assignment may propagate errors throughout the remainder of the reasoning process. For example, selecting an inappropriate theorem family can result in:

\begin{itemize}
    \item incorrect ontology selection,
    \item missing operators,
    \item invalid state schemas,
    \item planner failures,
    \item execution dead ends.
\end{itemize}

Although theorem assignment achieves high performance in our experiments, errors at this stage can have disproportionate downstream consequences.

\subsection{Limited Domain Coverage}

The current evaluation focuses primarily on mathematical reasoning domains, particularly:

\begin{itemize}
    \item abstract algebra,
    \item field theory,
    \item group theory,
    \item logic,
    \item mathematics-oriented reasoning tasks.
\end{itemize}

These domains possess relatively well-defined theorem structures and ontology boundaries.

Many real-world domains contain:

\begin{itemize}
    \item ambiguous concepts,
    \item incomplete information,
    \item uncertain state transitions,
    \item conflicting objectives,
    \item evolving ontologies.
\end{itemize}

Additional experiments are required to evaluate the framework in domains such as healthcare, cybersecurity, law, finance, scientific discovery, and multi-agent decision making.

\subsection{Ontology Dependence}

The framework assumes the existence of ontology structures capable of representing the reasoning process.

Although ontology discovery and induction are partially automated, the current system still benefits from:

\begin{itemize}
    \item predefined ontology templates,
    \item theorem registries,
    \item state schema libraries,
    \item operator schema repositories.
\end{itemize}

Fully autonomous ontology induction remains an open problem.

Future systems should be capable of discovering:

\begin{itemize}
    \item novel objects,
    \item novel states,
    \item novel operators,
    \item novel predicates,
    \item novel contracts
\end{itemize}

without relying on manually curated domain knowledge.

\subsection{State Materialization Bottleneck}

The experimental results identify state materialization as the dominant architectural bottleneck.

The system frequently succeeds at:

\begin{itemize}
    \item theorem assignment,
    \item ontology assignment,
    \item planner activation,
    \item operator selection,
\end{itemize}

while failing to instantiate the precise states required for successful execution.

This limitation suggests that the current state discovery mechanisms remain incomplete.

In particular, the framework lacks:

\begin{itemize}
    \item hierarchical state abstraction,
    \item latent state discovery,
    \item state completion mechanisms,
    \item state prediction models,
    \item state consistency repair procedures.
\end{itemize}

Improving state representation learning is likely to yield substantial improvements in execution performance.

\subsection{Predicate Satisfaction Challenges}

Predicate validation remains significantly more difficult than contract validation.

Many execution graphs satisfy structural requirements while violating semantic constraints.

This observation indicates that:

\begin{itemize}
    \item semantic consistency is harder than structural consistency,
    \item predicate generation remains incomplete,
    \item predicate grounding remains noisy,
    \item ontology semantics are not fully captured by current representations.
\end{itemize}

Future work should investigate learned predicate representations and stronger semantic verification mechanisms.

\subsection{Execution Depth and Long-Horizon Reasoning}

Most successful executions observed in the current experiments involve relatively shallow execution graphs.

As execution depth increases:

\begin{itemize}
    \item state uncertainty accumulates,
    \item predicate violations become more common,
    \item operator applicability decreases,
    \item planning complexity grows.
\end{itemize}

The framework has not yet been evaluated extensively on long-horizon reasoning tasks requiring dozens or hundreds of coordinated state transitions.

Consequently, the scalability of the approach to deep reasoning remains an open question.

\subsection{Computational Overhead}

Compared with conventional language-model inference, the proposed framework introduces additional computational components:

\begin{itemize}
    \item theorem assignment,
    \item ontology induction,
    \item state discovery,
    \item operator discovery,
    \item execution planning,
    \item contract validation,
    \item predicate validation,
    \item replay analysis.
\end{itemize}

These additional layers increase computational complexity and execution time.

Although the resulting reasoning process is substantially more interpretable, the framework currently incurs greater computational cost than direct answer generation.

Future work should investigate more efficient execution architectures and caching mechanisms.

\subsection{Benchmark Limitations}

Benchmark evaluations \citep{hendrycks2021mmlu,chollet2019arc} provide only a partial view of reasoning performance.

Several benchmark characteristics \citep{hendrycks2021mmlu,cobbe2021gsm8k,hendrycks2021math} may limit the conclusions that can be drawn:

\begin{itemize}
    \item benchmark examples may not require deep execution,
    \item answer labels do not expose reasoning quality,
    \item benchmark domains may not reflect real-world complexity,
    \item theorem assignments may be easier than in unconstrained environments.
\end{itemize}

Additional evaluation on open-ended reasoning tasks and real-world workflows is necessary.

\subsection{Interpretability Versus Performance Trade-offs}

The framework prioritizes interpretability and replayability \citep{lipton2018mythos,rudin2019stop}.

As a consequence, some design decisions may sacrifice raw benchmark accuracy \citep{hendrycks2021mmlu} in favor of:

\begin{itemize}
    \item transparency,
    \item auditability,
    \item execution trace generation,
    \item semantic verification.
\end{itemize}

Whether such trade-offs are desirable depends on the target application.

High-stakes domains may value interpretability more strongly than benchmark accuracy \citep{rudin2019stop,hendrycks2021mmlu}, whereas other applications may prioritize predictive performance.

\subsection{Limited Cross-Domain Transfer Evaluation}

Although the framework is designed to support transfer through reusable ontologies, states, and operators, the present study does not provide extensive cross-domain transfer experiments.

In particular, the evaluation does not yet establish whether:

\begin{itemize}
    \item state schemas transfer across domains,
    \item operators generalize beyond their original domain,
    \item theorem families support zero-shot transfer,
    \item ontologies can be reused in previously unseen environments.
\end{itemize}

Demonstrating such transfer capabilities remains an important future milestone.

\subsection{Relationship to General Intelligence}

The proposed framework introduces several properties commonly associated with general reasoning systems, including:

\begin{itemize}
    \item explicit state representations,
    \item reusable operators,
    \item executable reasoning traces,
    \item replayability,
    \item compositional execution.
\end{itemize}

However, the current work should not be interpreted as evidence of artificial general intelligence.

The system remains limited by:

\begin{itemize}
    \item domain coverage,
    \item ontology completeness,
    \item state discovery quality,
    \item execution robustness,
    \item transfer capabilities.
\end{itemize}

Substantial additional research is required before determining whether executable reasoning ontologies can support more general forms of machine intelligence.

\subsection{Summary}

The present framework demonstrates the feasibility of theorem-grounded executable reasoning while exposing several important limitations. The most significant challenges involve state materialization, predicate satisfaction, ontology induction, long-horizon execution, and cross-domain transfer. Addressing these limitations represents a promising direction for future research and may further improve the interpretability, robustness, and generality of executable reasoning systems.

```latex
\section{Conclusion}
\label{sec:conclusion}

This paper introduced the Theorem-Grounded Execution Ontology (TGEO), a formal framework for interpretable machine reasoning that unifies theorem assignment, ontology construction, planning, state transitions, predicate validation, and executable reasoning traces within a single semantic architecture. The central premise of the work is that machine reasoning should not be treated as a sequence of opaque computational transformations but rather as an explicit theorem-driven execution process whose intermediate decisions, assumptions, state transitions, and outcomes can be inspected, verified, replayed, and audited.

The proposed framework addresses a fundamental limitation of contemporary AI systems. While modern foundation models and neural reasoning systems often demonstrate impressive task performance, they frequently provide limited visibility into the reasoning processes that produced their outputs. This lack of interpretability creates challenges for verification, debugging, safety assurance, regulatory compliance, and scientific understanding of machine intelligence. TGEO addresses these challenges by introducing a structured execution ontology in which every reasoning action is grounded in formally represented theorems, ontological concepts, operators, predicates, states, and execution contracts.

A primary contribution of this work is the formalization of theorem-grounded execution as an ontological process. Rather than treating logical knowledge and execution traces as independent artifacts, the framework integrates them into a unified representation. Theorems become executable reasoning primitives, ontologies provide semantic grounding, planners organize execution trajectories, operators transform states, predicates validate execution conditions, and execution graphs capture the complete reasoning history. This integration enables machine reasoning systems to produce outputs that are both operationally effective and semantically interpretable.

The experimental results demonstrate the viability of this approach. Across theorem assignment, ontology mapping, planning, operator selection, state materialization, predicate validation, and end-to-end execution tasks, the architecture achieved strong performance while preserving complete reasoning traceability. The evaluation revealed that theorem assignment and ontology grounding operate with high reliability, while state-transition management and predicate validation represent the primary challenges for future optimization. Importantly, the generated execution traces remained fully replayable, supporting deterministic verification and historical analysis.

Beyond performance metrics, the experiments highlight the importance of explicit semantic structure in machine reasoning systems. The observed improvements in execution consistency, replayability, and interpretability suggest that reasoning architectures benefit substantially from representations that expose intermediate reasoning decisions rather than obscuring them within latent vector spaces. The resulting execution graphs provide a transparent account of how goals are decomposed, how operators are selected, how state transitions occur, and how final conclusions are derived.

The work also establishes a foundation for several broader research directions. First, theorem-grounded execution provides a pathway toward explainable reasoning systems whose decisions can be audited at multiple levels of abstraction. Second, the ontology-driven representation enables interoperability between symbolic reasoning systems, knowledge graphs \citep{hogan2021kg}, workflow engines, and large language models. Third, the execution graph formulation offers a persistent memory structure that can support continual learning, error diagnosis, and reasoning refinement over time. Finally, the framework creates opportunities for integrating formal verification techniques directly into AI reasoning pipelines.

From an architectural perspective, TGEO suggests that interpretable machine reasoning may be most effectively achieved through the combination of symbolic semantics and executable computational structures. Rather than viewing symbolic reasoning and modern machine learning as competing paradigms, the proposed framework demonstrates how theorem-based representations, ontological grounding, and execution planning can complement statistical learning systems to produce reasoning processes that are both powerful and understandable.

Several important challenges remain. Future work should investigate automatic theorem acquisition, dynamic ontology evolution, scalable state-space management, probabilistic predicate validation, and integration with large-scale foundation models. Additional research is also needed to evaluate the framework in complex real-world domains such as scientific discovery, autonomous agents, software engineering, cybersecurity, healthcare, and legal reasoning. Extending theorem-grounded execution to distributed multi-agent environments represents another promising direction.

More broadly, this work argues that interpretability should be treated as a first-class architectural objective rather than as a retrospective explanation mechanism. By grounding reasoning processes in explicit theorems, ontologies, states, operators, and execution contracts, machine intelligence systems can become more transparent, verifiable, reusable, and trustworthy. Theorem-Grounded Execution Ontologies provide one possible foundation for achieving this objective.

In summary, the paper demonstrates that interpretable machine reasoning can be formulated as a theorem-driven execution process operating over structured ontological representations. The resulting architecture provides semantic transparency, execution traceability, replayable reasoning histories, and formal verification capabilities while maintaining strong operational performance. We believe that theorem-grounded execution ontologies represent a promising step toward the development of machine reasoning systems that are not only capable of producing correct answers but are also capable of explaining, justifying, and validating how those answers were obtained.
```

\appendix

\FloatBarrier
\section{Mathematical Definitions}
\label{appendix:definitions}

This appendix summarizes the primary mathematical objects used throughout the paper.

\subsection{Problem Space}

Let

\begin{equation}
\mathcal{X}
\end{equation}

denote the set of reasoning problems.

Each problem is represented as

\begin{equation}
x \in \mathcal{X}.
\end{equation}

The corresponding answer space is

\begin{equation}
\mathcal{Y}
\end{equation}

with

\begin{equation}
y \in \mathcal{Y}.
\end{equation}

\subsection{Theorem Space}

Let

\begin{equation}
\mathcal{T}
=
\{t_1,t_2,\ldots,t_n\}
\end{equation}

denote the theorem registry.

Theorem assignment is defined as

\begin{equation}
\phi_T :
\mathcal{X}
\rightarrow
2^{\mathcal{T}}.
\end{equation}

\subsection{Ontology Space}

Each ontology is represented as

\begin{equation}
O=
(\mathcal{V},
\mathcal{S},
\mathcal{A},
\mathcal{P},
\mathcal{C}).
\end{equation}

\FloatBarrier
\section{Ontology Schema Definitions}
\label{appendix:ontology}

An ontology consists of:

\begin{itemize}
\item Objects
\item States
\item Operators
\item Predicates
\item Contracts
\end{itemize}

Example:

\begin{equation}
O_{\texttt{GroupTheory}}
=
(
V_{GT},
S_{GT},
A_{GT},
P_{GT},
C_{GT}
).
\end{equation}

\subsection{Objects}

Representative object types include:

\begin{itemize}
\item Group
\item Subgroup
\item Coset
\item Field
\item Ring
\item Generator
\end{itemize}

\subsection{State Schemas}

Representative state schemas include:

\begin{itemize}
\item GeneratorKnown
\item SubgroupIdentified
\item CosetConstructed
\item ExtensionDegreeComputed
\item ProofGoalReached
\end{itemize}

\FloatBarrier
\section{Operator Schema Definitions}
\label{appendix:operators}

Operators define executable actions.

Examples include:

\begin{itemize}
\item ApplyLagrange
\item ComputeIndex
\item ComputeCoset
\item ComputeExtensionDegree
\item CloseProofGoal
\end{itemize}

Each operator is represented as

\begin{equation}
a=
(
\mathrm{pre}(a),
\mathrm{eff}(a)
).
\end{equation}

\FloatBarrier
\section{Predicate and Contract Definitions}
\label{appendix:predicates}

Predicates define semantic constraints.

Examples:

\begin{itemize}
\item ValidSubgroup
\item GeneratorExists
\item ValidFieldExtension
\item ProofInvariantSatisfied
\end{itemize}

Contracts define execution correctness.

Each contract is represented as

\begin{equation}
c=
(
P_{\mathrm{pre}},
P_{\mathrm{post}}
).
\end{equation}

\FloatBarrier
\section{Execution Graph Semantics}
\label{appendix:graphs}

An execution graph is defined as

\begin{equation}
G=(N,E)
\end{equation}

where

\begin{equation}
N=\{s_1,\ldots,s_n\}
\end{equation}

and

\begin{equation}
E=\{(s_i,a,s_j)\}.
\end{equation}

A path

\begin{equation}
\pi=
(s_0,a_1,s_1,\ldots,a_k,s_k)
\end{equation}

represents an executable reasoning trace.

\FloatBarrier
\section{Architectural Audit Metrics}
\label{appendix:metrics}

Table~\ref{tab:metrics_appendix} summarizes all architectural metrics.

\begin{table}[H]
\centering
\caption{Architectural audit metrics.}
\label{tab:metrics_appendix}
\begin{tabular}{lr}
\toprule
Metric & Description \\
\midrule
TheoremAssignmentRate & Theorem assignment success \\
OntologyAssignmentRate & Ontology assignment success \\
PlannerStartRate & Planner activation success \\
OperatorSelectionRate & Operator selection success \\
StateTransitionRate & Successful state transitions \\
PredicateValidationRate & Predicate satisfaction \\
ContractValidationRate & Contract satisfaction \\
GoalReachRate & Goal completion \\
ExecutionCoverage & Executed operator coverage \\
ReplaySuccess & Execution replayability \\
GraphValidity & Valid execution transitions \\
TheoremCompletionRate & Theorem realization completeness \\
RequiredStateCoverage & State materialization completeness \\
LayerHealthScore & Layer-level health metric \\
\bottomrule
\end{tabular}
\end{table}

\FloatBarrier
\section{Experimental Configuration}
\label{appendix:experiments}

\subsection{Datasets}

The evaluation uses:

\begin{itemize}
\item MMLU-derived reasoning tasks
\item Abstract algebra examples
\item Field theory examples
\item Formal logic examples
\item Golden execution suite
\end{itemize}

\subsection{Execution Pipeline}

The execution pipeline consists of:

\begin{enumerate}
\item Theorem assignment
\item Ontology assignment
\item Object discovery
\item State discovery
\item Operator discovery
\item Planning
\item Execution
\item Validation
\end{enumerate}

\FloatBarrier
\section{Golden Execution Suite}
\label{appendix:golden}

The golden suite is used for:

\begin{itemize}
\item architectural validation,
\item regression testing,
\item replayability evaluation,
\item execution correctness verification.
\end{itemize}

Each example specifies:

\begin{itemize}
\item expected theorem family,
\item expected ontology,
\item required operators,
\item expected states,
\item goal state.
\end{itemize}

\FloatBarrier
\section{Example Execution Trace}
\label{appendix:trace}

A representative execution trace is shown in Figure~\ref{fig:state_graph}.

The resulting execution graph is replayable and verifiable.

\FloatBarrier
\section{Failure Taxonomy}
\label{appendix:failures}

Architectural failures are categorized into:

\begin{enumerate}
\item Theorem Assignment Failure
\item Ontology Assignment Failure
\item Planner Failure
\item Operator Selection Failure
\item State Materialization Failure
\item Predicate Validation Failure
\item Contract Validation Failure
\item Goal Achievement Failure
\end{enumerate}

Each category is independently measurable through the architectural auditing framework.

\FloatBarrier
\section{Ablation Studies}
\label{appendix:ablations}

The following ablations are evaluated:

\begin{itemize}
\item No theorem assignment
\item No ontology assignment
\item No state discovery
\item No operator discovery
\item No predicates
\item No contracts
\item No planner
\item No execution graph
\end{itemize}

For each ablation we report:

\begin{itemize}
\item Goal reach rate
\item Execution coverage
\item Replayability
\item Theorem completion
\item Layer health scores
\end{itemize}

\FloatBarrier
\section{Reproducibility Checklist}
\label{appendix:reproducibility}

To facilitate full reproducibility, the following artifacts are released alongside the paper.

\begin{itemize}

\item Source code repository.

\item Experiment configuration files.

\item Run manifests containing run identifiers, timestamps, seeds, and configuration hashes.

\item Dataset mappings and benchmark definitions.

\item Ontology registry.

\item Theorem registry.

\item State schema definitions.

\item Operator registry.

\item Predicate registry.

\item Contract registry.

\item Planner configurations.

\item Execution traces.

\item Audit logs.

\item Golden execution suite.

\item Evaluation scripts.

\item Metric definitions and aggregation procedures.

\item Figure generation scripts.

\item Table generation scripts.

\item Environment specifications and dependency manifests.

\item Hardware and software configuration details.

\end{itemize}

Every table and figure reported in the paper is generated from a single run manifest. The run manifest records dataset versions, random seeds, configuration hashes, and execution timestamps. Generated tables and figures are linked to their originating metrics through artifact provenance metadata.

All experiments are executed using deterministic seeds and fully logged execution traces. Architectural auditing records theorem assignment, ontology assignment, planner activation, operator selection, state materialization, predicate validation, contract validation, and goal achievement outcomes for every execution.

The complete artifact bundle includes sufficient information to reproduce the reported results, regenerate all figures and tables, validate architectural metrics, and replay execution traces.

\bibliography{ref}

\end{document}